\documentclass[sn-mathphys]{sn-jnl}

\jyear{2021}%

\theoremstyle{thmstyleone}%
\theoremstyle{thmstyletwo}%

\theoremstyle{thmstylethree}%

\begin{document}


\title[The bulk flow motion and the Hubble-Lema\^{\i}tre law in the Local Universe]
{The bulk flow motion and the Hubble-Lema\^{\i}tre law in the Local Universe with the ALFALFA survey}


\author*[1]{\fnm{Felipe} \sur{Avila}}\email{felipeavila@on.br}

\author[2]{\fnm{Jezebel} \sur{Oliveira}}\email{jezebel21@ov.ufrj.br}
\equalcont{These authors contributed equally to this work.}

\author[3]{\fnm{Mariana} \sur{L. S. Dias}}\email{lopesmariana@id.uff.br}
\equalcont{These authors contributed equally to this work.}

\author[1]{\fnm{Armando} \sur{Bernui}}\email{bernui@on.br}
\equalcont{These authors contributed equally to this work.}

\affil*[1]{Observat\'orio Nacional, Rua General Jos\'e Cristino 77, 
	S\~ao Crist\'ov\~ao, 20921-400 Rio de Janeiro, RJ, Brazil}

\affil[2]{Observat\'orio do Valongo, Ladeira do Pedro Ant\^onio, 43 - Centro, 
	20080-090 Rio de Janeiro, RJ, Brazil}

\affil[3]{Universidade Federal Fluminense, Rua Passo da P\'atria, 152, S\~ao Domingos, 24210-240 Niter\'oi - RJ, Brazil}


\abstract{
The knowledge of the main features of the bulk flow in the Local 
Universe is important for a better determination of the relative motions there, 
an information that would contribute to a precise calculation of the 
Hubble-Lema\^{\i}tre law at very low redshifts. 
We study how to obtain the Hubble-Lema\^{\i}tre law in two sky regions using 
the catalog of HI sources of the ALFALFA survey, with data 
$cz_{\odot} < 6000$ km/s. 
Our methodology aims to compute $H_0$ in two regions --located in opposite galactic 
hemispheres-- mapped by the ALFALFA survey, and look for dependence with distance, 
direction, and also test for reference frame changes. 
We calculate the Hubble constant, in the Cosmic Microwave Background reference frame, in opposite galactic hemispheres: $H_0^N = 70.87 \pm 2.38$ and 
$H_0^S = 66.07 \pm 3.02$, which allows us to measure the bulk flow velocity $V_{\mbox{\tiny BF}} = 401.06 \pm 150.55$ km/s at the effective distance 
$31.3 \pm 6.26$ Mpc, a novel result found analysing the ALFALFA data at low redshift. 
We confirm the influence of the bulk flow on the structures of the Local Universe which manifests through a dipolar behavior of the Hubble constant in opposite hemispheres.}




\maketitle

\section{Introduction}\label{introduction}
The peculiar motions of the galaxies are a consequence of the gravitational field originated by the matter 
distribution~\cite{Peebles80}.  
In particular, the presence in the Local Universe of under-dense regions, like the large region 
void of galaxies called the {\em Local Void}~\cite{Tully88,Tully08}, 
and over-dense volumes hosting clusters and superclusters of galaxies, like Virgo, Hercules, 
Coma, Pisces-Perseus, etc.~\cite{Jerjen93,Courteau99}, 
substantially dictates the dynamical features of nearby galaxies opening a competition 
between the gravitational field attraction against the cosmic Hubble expansion~\cite{Lavaux10,Bilicki11,Hoffman17}. 
According to the Hubble-Lema\^{\i}tre law, the expansion rate is small at small distances from us 
and in such a case peculiar velocities of nearby galaxies --pointing along arbitrary directions-- 
compete with the local Hubble flow --in radial direction-- affecting our measurement of their radial 
recessional velocities. 
In fact, when observing a galaxy to measure its recessional velocity what we are actually measuring is 
the resultant of the vector sum of its radial Hubble speed plus its peculiar velocity. 
For this, one does not expect that the galaxies near us, i.e., $z < 0.02$, follow the Hubble flow 
and recess from us at an equal rate as distant galaxies, i.e., $z \gg 0.02$, do. 
As a consequence, our local cosmic web with over- and under-dense regions sensibly affects the final motion 
of the galaxies in the Local Universe, preventing them to exactly follow the Hubble 
flow~\cite{Mendez02,Karachentsev02,Karachentsev07,Karachentsev09}. 
For this, the knowledge of the main features of the bulk flow motion could be important to quantify the magnitude 
of the peculiar motions, a necessary knowledge for a correct  calculation of the Hubble-Lema\^{\i}tre law. 
The study of large-scale matter distribution is not limited to the Local Universe, current investigations 
encompass diverse clustering analyses at several redshift ranges of astronomical data surveys 
(see, e.g., \cite{Bengaly17,Sarkar19,Marques20a,Marques20b,Dong21,Edilson21,Avila19,Avila22a,Perenon22}).

To the traditional interest in the light-curves of SNIa to measure cosmological distances, there are 
recent efforts to improve the calibration methods of other observables like Cepheids and of 
Tip of the Red Giant Branch (TRGB)~\cite{Freedman19,Yuan19} to be employed as standard candles. 
Because of the tension found in the values of the Hubble constant, $H_0$, calculated with 
distinct cosmological probes in the late~\cite{Riess19}
and early universe~\cite{Planck20}, 
this panorama is now changing. 
Nowadays, there is a growing interest in examining diverse cosmological probes that can be used to measure the Hubble constant --and other cosmological parameters-- and in the statistical analyses that can be done with them~(see, e.g.,~\cite{Benisty21,Benisty22,Staicova22,Avila22b,Nunes20,Nunes21,
Yang22,Alestas22,Mokeddem22,Perivolaropoulos22,Valentino22a,Valentino22b}).

\begin{figure}
	\begin{center}
		\vspace{-0.5cm}
		\includegraphics[width=9.1cm,height=6.0cm]{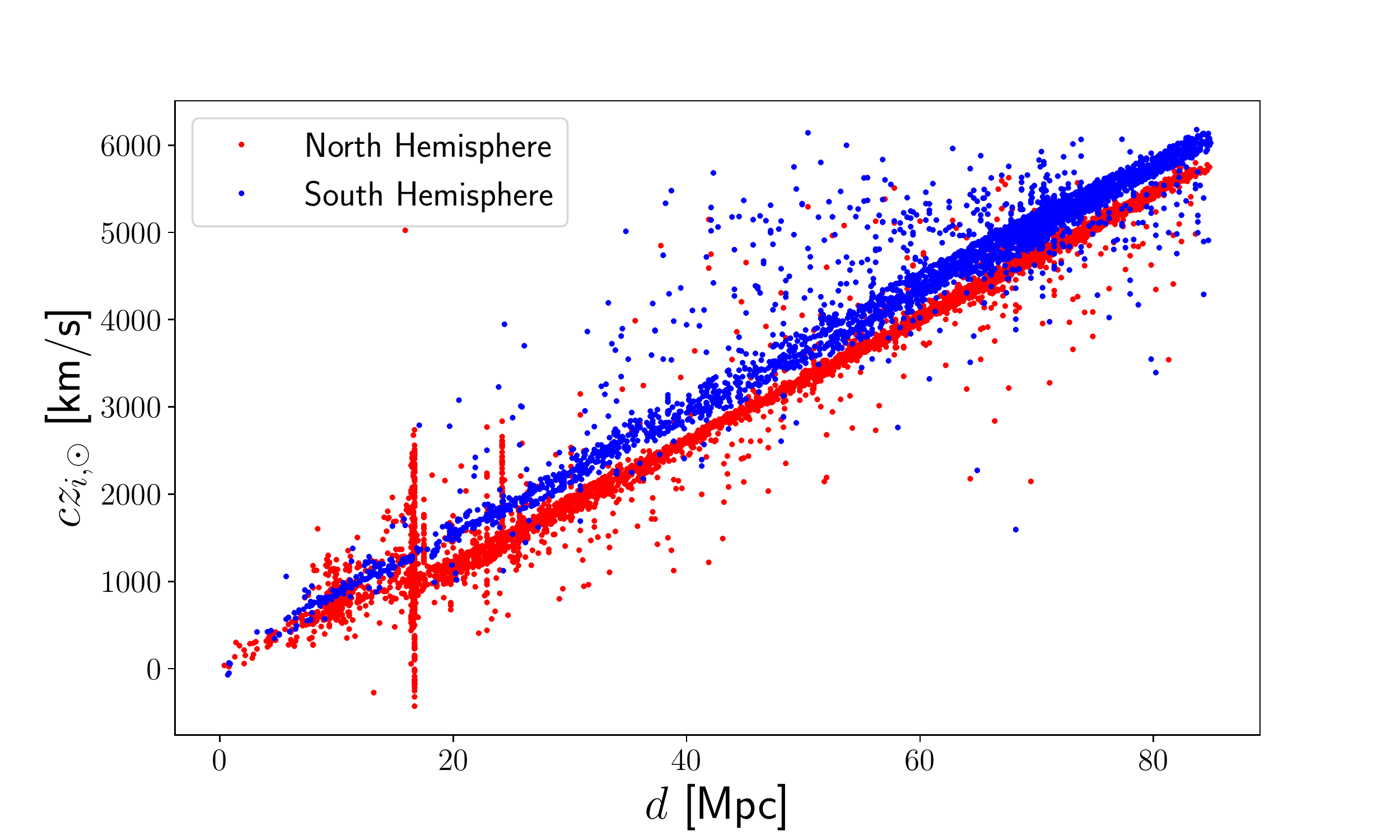}
	\end{center}
	\vspace{0.cm}
	\caption{Hubble-Lema\^{\i}tre diagram of the North (red points, originally termed Spring) and South 
		(blue points, originally termed Fall) data sets regions observed in the ALFALFA survey. 
		As the velocities of these extra-galactic objects are in the heliocentric reference frame, we observe a offset 
		between hemispheres, indicating a peculiar motion of the solar system with respect to large scale structures, 
		which prevents directly calculating the Hubble constant.}
	\label{fig1}
\end{figure}

Peculiar motions can be revealed in any Hubble-Lema\^{\i}tre diagram through the scatter of the 
data above and below the linear correlation expected, due to the universe expansion, between 
the distance and velocity of each cosmic object. 
In fact, in Figure~\ref{fig1} we observe the Hubble-Lema\^{\i}tre diagram of a data set of extra-galactic 
HI sources with heliocentric velocities $c z_{\odot} < 6\,000$ km/s, a subset of the ALFALFA 
survey data \cite{Haynes18} whose distances were measured without using the Hubble law hypothesis, 
where one notices a large dispersion of data away from the expected linear distance--velocity 
relationship. 
These cosmic objects belong to the Local Universe, i.e. $z < 0.02$, and were observed in two 
disjoint patches of the celestial sphere termed {\em Spring} (located in the North galactic hemisphere) 
and {\em Fall} (located in the South galactic hemisphere) by the ALFALFA collaboration \cite{Haynes18}. 
The Hubble-Lema\^{\i}tre diagram shown in Figure~\ref{fig1} also evidences a second phenomenon: 
a distinct recession velocity for the data in each sky patch. 
This anisotropic behavior indicates that the Sun's reference frame, given that the velocities in the ALFALFA 
catalog are heliocentric $c z_{\odot}$, is not suitable to analyse the Hubble-Lema\^{\i}tre diagram and the 
heliocentric velocity data needs a reference frame transformation. 
This fact is already expected due to gravitational attractors in the Local Universe that produce relative 
motions between the cosmic objects, a phenomenon that has been studied since the 
1980's \cite{Tammann85,Tully88,Jerjen93,Pike05,Kocevski06,Carrick15,Said20,Boruah20,Boruah21}.


We restrict our study of the ALFALFA catalogue to cosmic objects with velocities $c z_{\odot} < 6\,000$ km/s 
because the distances for these data were measured without using the Hubble-Lema\^{\i}tre law, and for this 
they can be used to examine interesting questions like: 
how much the gravitational field of local structures affects the measurement of the Hubble constant in the 
Local Universe? 
This question motivate us to perform detailed analyses that takes into account velocity reference-frame 
transformations to calculate the Hubble constant $H_0$ in the Local Universe using the ALFALFA survey data. 
Here we shall investigate the Hubble flow for galaxies in the Local Universe, and for this we perform suitable 
reference frame transformations. 
For velocity data measured in the heliocentric frame one has to consider two velocity-frame transformations \cite{Erdogdu06}: 
(1) first, one considers the relative motion between the Sun with respect to the barycenter of the Local Group (LG) 
of galaxies; 
(2) second, for galaxies far from the LG, i.e. $r \gtrsim 10$ Mpc, it is observed a net motion towards the 
cosmic microwave background (CMB) dipole apex, for this the velocities referred to the LG of galaxies should be 
transformed to the CMB reference frame (this velocity transformation takes into account the gravitational 
attraction of the LG towards the Virgo cluster, phenomenon termed the Virgocentric infall, see e.g. \cite{Tammann12,Courteau99,Lineweaver95}). 

Another important systematic effect is the {\em cosmic variance}, also termed sample variance~\citep{Driver10}. 
Measurements of $H_0$ in small volumes, as in the Local Universe, will be largely 
affected by peculiar velocities as explained above. 
The literature reports that the cosmic variance contributes, in the computation of $H_0$, 
with $\lesssim 3\%$ for $z > 0.01$ and $\lesssim 2\%$ for $z > 0.02$~\citep{Marra13,Wu17,Camarena18}. 
While such an effect could be ruled out simply by performing a redshift cut on the sample in analysis, as in the case of supernovae data, we cannot, however, do the same here. 
But, as we shall discuss in section~\ref{error}, the distance errors of the ALFALFA 
galaxies (HI extra-galactic sources) are of the order of 20\%, which implies an error 
of the order of 5\% in $H_0$, which allows us to neglect the cosmic variance in our 
analyses, by theoretical and observational reasons~\cite{Marra13, Wu17, Camarena18}.

Additionally, the measurements of $H_0$ in opposite hemispheres, after the ALFALFA data was transformed to 
the CMB reference frame, confirm the influence of the bulk flow on the structures of the Local Universe, which manifests 
through a dipolar behavior of the Hubble constant in opposite hemispheres, a result that allows us to calculate the bulk 
flow velocity, a novel result not found in the literature using the ALFALFA data.

It becomes important to distinguish our analyses, which calculate the 
Hubble-Lema\^{\i}tre diagram and measure $H_0$, with others in the literature. 
We made this study to understand the effect of local structures on the measurement 
of $H_0$; 
thus our result can not be considered as actually a final result because, as our 
analyses show, the effect of the bulk flow is present and relevant. 
Therefore, our measurement of $H_0$ represents just the outcome of 
the statistical methodology applied here to investigate the dependence of the Hubble-Lema\^{\i}tre law on the physical features of the Local Universe.


%

\vspace{0.5cm}

This work is organized as follows: 
in section \ref{data} we describe the ALFALFA data and the sample selected for our analyses. 
In section \ref{method} we describe the procedures for the velocity frame transformations and 
the way how to obtain the best interval for calculating the Hubble constant $H_0$ (due to the dispersion 
of the data in the Hubble-Lema\^{\i}tre diagram). 
In section \ref{results} we show the details to calculate $H_0$, including the systematic errors. 
In section~\ref{conclusions} we discuss our results and present our conclusions. 
We leave the robustness tests of our results to the Appendices section.

\section{The Arecibo Legacy Fast ALFA Survey}\label{data}

The Arecibo Legacy Fast ALFA (ALFALFA) Survey
mapped extra-galactic HI line sources, in 21 cm, covering a sky region of 
$\sim \!\!7\,000 \,\text{deg}^{2}$. 
The completed survey produced a catalogue of $\sim \!31\,500$ extra-galactic HI line sources, 
mainly hydrogen gas-rich galaxies 
with low surface brightness and dwarfs that populates the local universe, 
$0 < z < 0.06$ (\cite{Giovanelli15,Haynes18}; see also \cite{Avila18,Avila21} for applications of these data).

The sources catalogued in this survey are classified in three categories according to its HI line 
detection features: \\
-- {\sc code} 1, high signal to noise ratio extra-galactic sources, considered highly reliable and with 
confirmed optical counterpart; \\
-- {\sc code} 2, lower signal to noise ratio HI signal coincident with optical counterpart, considered 
unreliable sources; and \\
-- {\sc code} 9, high signal to noise ratio source with no optical counterpart and likely Galactic high 
velocity cloud. \\
The ALFALFA team recommends using the HI sources designated with {\sc Code 1}.

\begin{figure}
	\begin{center}
		\includegraphics[width=8.0cm,height=5.0cm]{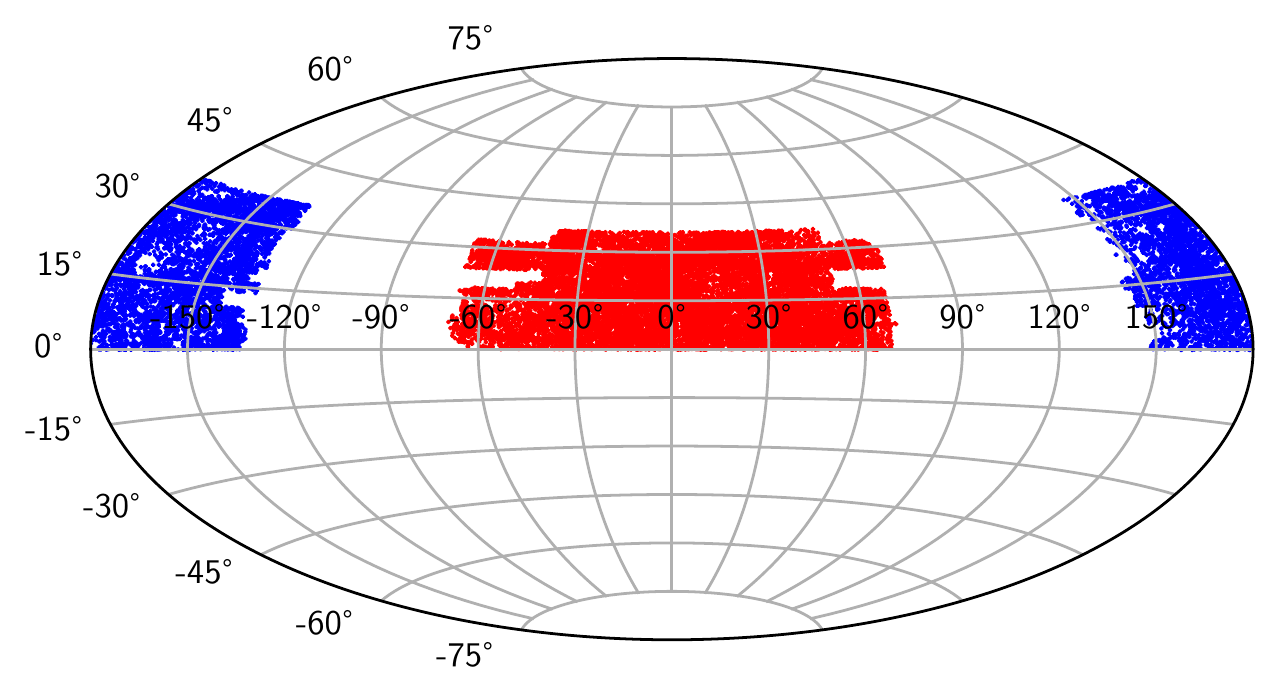}
	\end{center}
	\caption{Footprint of the ALFALFA survey, showing the observed sky regions, 
		the {\it Fall} or South (blue) and {\it Spring} or North (red).}
	\label{fig2}
\end{figure}

The ALFALFA survey covered two regions in the celestial sphere, with declination range 
$0^{\circ} < \text{DEC} < 36^{\circ}$, with right ascension intervals of 
$21^{\text{h}} 30^{\text{m}} < \text{RA} < 3^{\text{h}} 15^{\text{m}}$ and 
$7^{\text{h}} 20^{\text{m}} < \text{RA} < 16^{\text{h}}40^{\text{m}}$, 
as illustrated in the Figure~\ref{fig2}.


\subsection{Data selection}\label{sec2.1}

The heliocentric velocities in the ALFALFA survey were obtained, with very small uncertainties, 
through the measured spectra of the extra-galactic HI line sources.  
The distance measurements in the ALFALFA catalogue are described in Haynes et al. 2018 (see Section 3.1, column 11). 
The ALFALFA collaboration uses two distances estimation approaches: \\
(i) for those objects with $cz_{\odot} > 6\,000$ km/s the distance is simply estimated as 
$c z_{\mbox{\tiny CMB}}/H_0$, that is, through the Hubble-Lema\^{\i}tre law, where $c z_{\mbox{\tiny CMB}}$ is the 
recessional velocity measured in the Cosmic Microwave Background (CMB) radiation reference frame and 
$H_0$ is the Hubble constant; and, 
(ii) for objects with $c z_{\odot} < 6\,000$ km/s the collaboration assigns distances to galaxies 
using the local peculiar velocity model developed by \cite{Masters05} based mainly on the SFI++ catalogue 
of galaxies \cite{Springob07} and results from analyses of the peculiar motion of galaxies, groups, 
and clusters, using a combination of primary distances from the literature and secondary distances from 
the Tully-Fisher approach \cite{Tully13}.

Our methodology aims to compute $H_0$ in two sky regions –in opposite galactic hemispheres– mapped by the ALFALFA survey and look for dependence with direction, with distance, and
also test for effect of reference frame changes. 
These results will allow us to measure the bulk flow velocity, as we shall explain in section~\ref{bulkflow}.
To do this, we first analyze cosmic objects that have their distance measurements obtained without using the Hubble-Lema\^{\i}tre law. 
Therefore, we select galaxies with threshold velocity $6\,000$ km/s, but for this velocity the galaxies 
of the catalogue have distances scattered around $85 - 90$ Mpc; 
for definiteness we remove from the catalogue the HI sources with distances above $85$ Mpc. 
After these cuts, we have for analyses a sub-sample of the ALFALFA catalogue of HI sources with 
$c z_{\odot} < 6\,000$ km/s and distances less than $85$ Mpc. 
In this sample the North region, also referred as {\it Spring}, has $4\,602$ HI sources, and 
the South region, also referred as {\it Fall}, contains $3\,422$ HI sources. 
Our analyses aim to measure the Hubble constant in both regions after performing the velocity-frame 
transformations due to relative motions in the Local Universe \cite{Erdogdu06}. 

\section{Methodology}\label{method}

In this section we present the methodology to estimate the Hubble parameter, $H(r)$, 
for the selected sample from the ALFALFA catalogue. 
Notice that the velocity reference frame of the ALFALFA catalogue is heliocentric. 
According to the literature, our analyses will be done in two steps: first we consider the Local Group (LG) reference 
frame and then the CMB reference frame (see, e.g., \cite{Erdogdu06}).

After these reference frame transformations one must determine the best interval to implement our analyses. 
This is necessary because, as mentioned in section~\ref{introduction}, in the Local Universe the peculiar velocities 
are dominant over the Hubble flow, and one needs to know the scale $r_{0}$, which is the shortest distance where the 
peculiar velocities decrease significantly and the Hubble flow dominates. 
Then we shall show the statistical analyses used to calculate the Hubble constant and its corresponding uncertainty. 
After these analyses one find the Hubble constant in opposite hemispheres, 
and at the end these information is used to determine the bulk flow velocity, as we shall explain.

\subsection{Transforming the heliocentric frame of reference}\label{referenceframe}

We start our analyses by transforming the heliocentric velocities of the selected ALFALFA data sample, with 
$c z_{\odot} \le 6\,000$ km/s and $d \le 85$ Mpc, into the CMB reference frame, 
for this one has to consider two velocity-frame transformations: 
firstly, one transforms the heliocentric frame to the LG reference frame, then one changes the data 
velocities from the LG reference frame to the CMB reference frame.

\begin{figure}[h]
	\centering
	\includegraphics[scale=0.6]{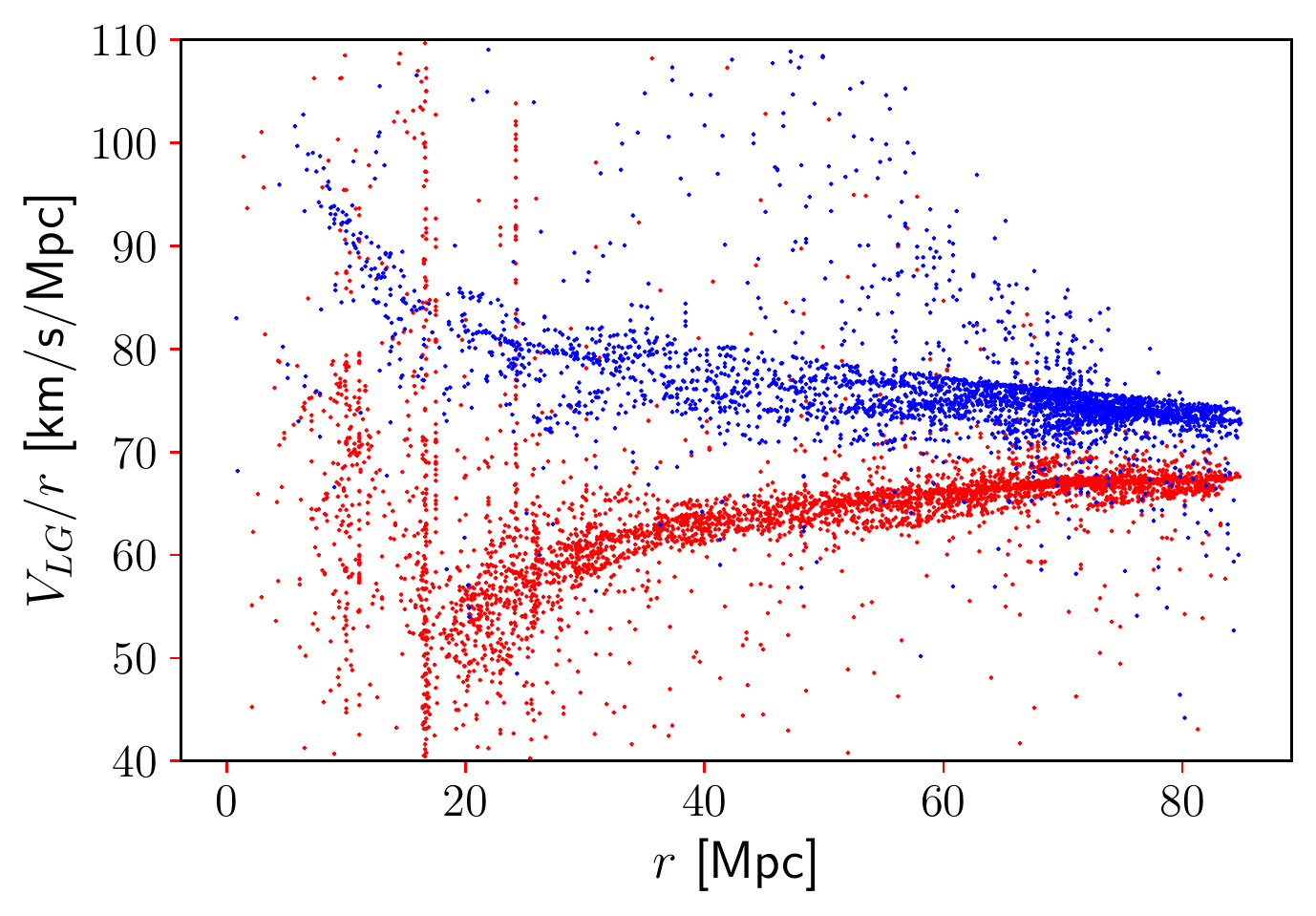}
	\caption{Plot of the ALFALFA data, from the North and South hemispheres, in the form: 
		$V_{\mbox{\sc lg}}/r$ {\it versus} $r$. 
	}
	\label{fig:Hubble_parameter_LG_frame}
\end{figure}

Our first step is to transform the heliocentric velocities of the ALFALFA catalogue to the barycenter 
of our LG of galaxies, to which the Milky Way is gravitationally bounded. 
One knows the Sun's velocity, direction and magnitude, with respect to the LG barycenter: 
$\mbox{\bf V}_{\odot/\mbox{\sc lg}} = 318 \pm 20$ km/s towards 
$(l, b) \,=\, (106^{\circ} \pm 4^{\circ}, - 6^{\circ} \pm 4^{\circ})$, a result based on the well 
determined motion of our Galaxy (Tully et al. 2008). 
Then, following \cite{Erdogdu06}, one transforms the heliocentric velocities of the ALFALFA catalogue, 
$c z_{\odot}$, to the barycenter of the LG frame using 
\begin{equation}\label{helio-LG}
	V_{i,\mbox{\sc lg}} = c z_{i,\odot} - 79 \cos l_i \cos b_i + 296 \sin l_i \cos b_i - 36 \sin b_i \,,
\end{equation}
where $l_i$ and $b_i$ are the galactic coordinates of the $i$-th galaxy and $z_{i,\odot}$ is its 
heliocentric redshift. 
In Figure~\ref{fig:Hubble_parameter_LG_frame} we observe the Hubble parameter, $V_{\text{LG}} /r$, 
for the LG reference frame, where the red and blue points correspond to data from the North and South hemispheres, respectively. 
One can notice a tendency towards a convergence for the data from both hemispheres. 
This figure motivates the analysis to find the suitable scale to 
calculate the Hubble constant.

\begin{figure}[h]
	\centering
	\includegraphics[scale=0.6]{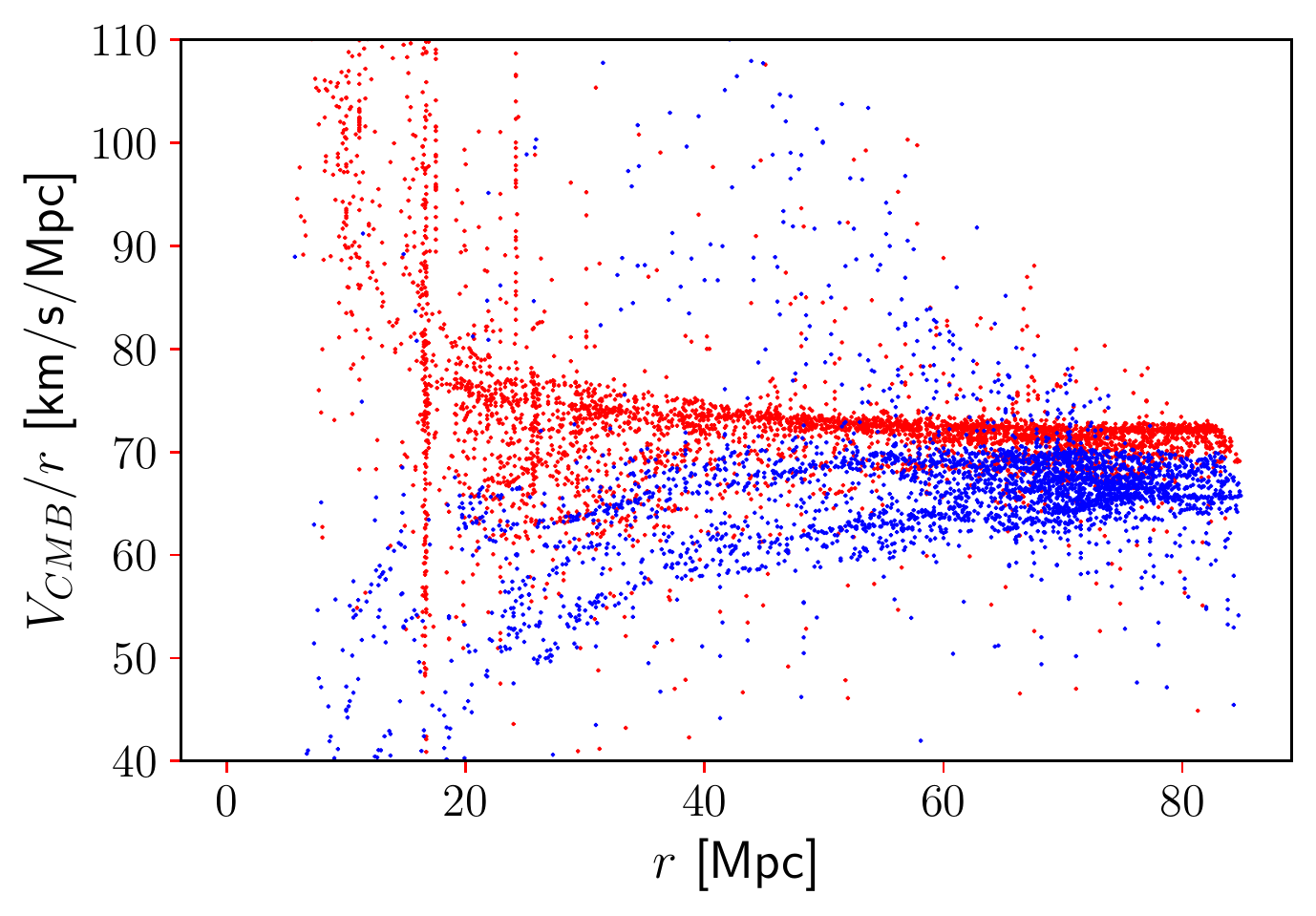}
	\caption{The Hubble parameter, plotted in the form $V_{\text{CMB}} /r$ {\it versus} $r$, 
		for data after the transformation to the second reference frame, i.e. the CMB reference frame, where the red 
		and blue points represent data from the North and South regions, respectively. In comparison with the LG frame, 
		we observe an approximation between the data from both hemispheres, and a convergence for scales above 40 Mpc.}
	\label{fig:Hubble_parameter_CMB_frame}
\end{figure}

The next step is to transform the velocities from the LG reference frame to the CMB frame using \cite{Courteau99,Erdogdu06} 
\begin{equation}\label{LG-CMB}
	V_{i, \text{CMB}} = V_{i,\mbox{\sc lg}} +  \mbox{v}_{\text{\sc LG/CMB}}[\sin(b_i)\sin(b_\text{LG})+\cos(b_i)\cos(b_\text{LG})\cos(\arrowvert l_{\text{LG}}-l\arrowvert)] \,,
\end{equation}
for the $i$-th galaxy, where $\mbox{v}_{\text{\sc LG/CMB}} = 627 \pm 22$ km/s is the LG velocity with respect to 
the CMB and $(l_{\text{LG}}, b_{\text{LG}})=(273^\circ \pm 3^\circ, 29^\circ\pm3^\circ)$. 
In Figure~\ref{fig:Hubble_parameter_CMB_frame} we observe the Hubble parameter, plotted in the form 
$V_{\text{CMB}} /r$ {\it versus} $r$, for data after the transformation to the second reference frame, i.e. the 
CMB reference frame, where the red and blue points represent data from the North and South regions, respectively. 
In this figure one can observe some interesting features. 
Firstly, for both data sets the $H(r)$ values seem to approach one to another. 
Second, the convergence shown by the data from the North region (red points) is more evident; 
but for the South region (blue points) the data appear disperse. 
Due to these features observed both in the LG and in the CMB frames, we shall apply 
suitable statistical tools in the next section.

\subsection{The Skewness test}\label{skewness}

As we observe in Figures~\ref{fig:Hubble_parameter_LG_frame} and~\ref{fig:Hubble_parameter_CMB_frame}, 
the plots show the Hubble parameter of the data in analysis having a large dispersion for low $r$ and 
an asymptotic convergence behavior for large $r$. 
As mentioned above, this behavior is expected because for distances $r < 40$ Mpc the cosmic objects feel the 
gravitational influence of the neighboring galaxies, influence that competes in intensity with the Hubble flow.  
For the North region, in particular, the dispersion decreases significantly after 20 Mpc. 
This is clear from the presence of the Virgo cluster located at around 16 Mpc from us. 
For the South region, the data is --reasonably-- well behaved for the whole sample. 
Therefore, in principle, one could propose for this study that the analyses will be done for data located 
from 20 Mpc until 85 Mpc, which is the limit of the selected sample.

Let us perform now a more detailed scrutiny of the initial value of the distance interval for 
analyses. 
For certain interval, $r_0 \le r \le 85$ Mpc ($r_0 \ge 20$ Mpc), we expect that, on average, 
the distribution of the data in the $V/r$ {\it versus} $r$ plot is well distributed around the mean value, 
i.e. the $H_0$ value~\footnote{Notice that the $V/r$ {\it versus} $r$ plot can be considered as a 
	distribution of $H(r)$ values, where it resembles a Gaussian or {\it normal} distribution with defined 
	parameters: mean and standard deviation.}. 
Thus, it seems appropriate to analyse the {\it normality} or Gaussianity of this distribution for different 
values of $r_0$. 
For example, one can use the third standardized moment of a given distribution \cite{Agostino90} 
\begin{equation}\label{skew}
	\sqrt{\beta} \equiv \frac{m_3}{m_2^{3/2}},    
\end{equation}
where 
\begin{equation}
	m_k = \frac{1}{n}\sum_i (X_i - \overline{X})^k \,,
\end{equation}
and $\overline{X}$ is the mean sample, defined as:
\begin{equation}
	\overline{X} \equiv \frac{1}{n} \sum_i X_i \,.
\end{equation}
The quantity $\sqrt{\beta}$ is the {\it skewness} of the distribution in analysis. 
In general, a value $\sqrt{\beta}$ close to 0 is an indication of Gaussianity, or normality, of the data sample. 
However, for some distributions the value of $\sqrt{\beta}$ is small but not zero, being not a clear indication 
of normality.

For this, we use an alternative and more powerful test of skewness \cite{Agostino90}. 
First, one transforms the original distribution, with skewness $\sqrt{\beta}$, to a normal distribution 
with zero mean and unity variance \cite{Agostino70}. 
For a null hypothesis of normality, and the number of elements in the sample be $n \ge 8$, one has the 
following equations 
\begin{equation}
	Y = \sqrt{\beta}\left[\dfrac{(n+1)(n+3)}{6(n-2)}\right]^{1/2} \,,
\end{equation}
\begin{equation}
	\gamma = \dfrac{3(n^2+27n-70)(n+1)(n+3)}{(n-2)(n+5)(n+7)(n+9)} \,,
\end{equation}
\begin{equation}
	W^2 = -1 + \left[2(\gamma - 1)\right]^{1/2} \,,
\end{equation}
\begin{equation}
	\delta = \frac{1}{\sqrt{\ln W}} \,,
\end{equation}
and 
\begin{equation}
	\alpha = \left(\frac{2}{W^2-1}\right)^{1/2} \,.
\end{equation}
Therefore, the expression 
\begin{equation}\label{zscore}
	Z(\sqrt{\beta}) = \delta\,\ln\left[\frac{Y}{\alpha} + \left(\frac{Y^2}{\alpha^2}+1\right)^{1/2}\right] \,,
\end{equation}
provides a measurement of the skewness of an approximately normal distribution with zero mean and unity variance, 
under the null hypothesis (for details see the appendix~\ref{ApendiceA}).

Then, we will analyze our samples (North and South) for different distance intervals and, from the Skewness test, 
we will determine the best interval for analysis, the one with low value of $Z$ and a statistical 
significance level with $p$-value greater than 0.01. In appendix \ref{ApendiceA}, we perform a null test for the code~\footnote{\url{https://docs.scipy.org/doc/scipy/reference/generated/scipy.stats.skewtest.html}} 
used to calculate $Z$ and the $p$-value.

After determining the suitable interval for our analyses using the Skewness test, i.e. $[r_0, 85]$ Mpc, 
we can calculate the mean and standard deviation of the distribution in the selected interval,
\begin{equation}\label{mean}
	H_0 = \frac{1}{n}\sum_i \frac{V_{i,\text{ref}}}{r_i} \,,
\end{equation}
and
\begin{equation}\label{std}
	\sigma_{H_0} = \sqrt{\frac{1}{n}\sum_i \left(\frac{V_{i,\text{ref}}}{r_i} - H_0\right)^2} \,,
\end{equation}
where $n$ is the number of objects in the interval in analysis and `ref' refers to the frame of reference, LG or CMB. 
Also, we calculate the median and the mode (using integer values of $V_{i,\text{ref}} / r_i$) to compare with $H_0$. 
For a normal distribution, these three values must be close. 

As a matter of robustness, in the appendix \ref{ApendiceB} we test several intervals for different 
mean (arithmetic, geometric, and harmonic) formulas to obtain $H_0$. 
Our intention with this robustness test is to check, with the Skewness test, whether our result is biased by the 
choice of the interval.

\subsection{Distance dependence analysis}\label{functions}

From the Skewness test, we determine the interval where the distribution $\{ V_i/r_i \}$ 
is close to a normal distribution, which allowed us to  calculate the value of the Hubble constant $H_0$.
However, we need to check first if, for this specific interval, there is a non-negligible distance-dependence.

In this work we fit our sample distribution with two functions 
\begin{equation}\label{F1}
	{}^{1}H(r) = a_1 r + b_1 \,,
\end{equation}
and 
\begin{equation}\label{F2}
	{}^{2}H(r) = \frac{a_2}{r} + b_2,
\end{equation}
where the left-side superscripts `1' and `2' are used to identify the two functions used for the best-fit, 
and $a_1, a_2$ and $b_1, b_2$ are the parameters to be adjusted. 
After this procedure we calculate the mean of the functions  ${}^{1}H$ and ${}^{2}H$ 
\begin{equation}\label{F1average}
	{}^{1}\overline{H} = \frac{a_1}{2}\frac{r^2_f - r^2_0}{r_f-r_0} + b_1 \,,
\end{equation}
and 
\begin{equation}\label{F2average}
	{}^{2}\overline{H} =  \frac{a_2}{r_f-r_0}\ln\left(\frac{r_f}{r_0}\right) + b_2 \,,
\end{equation}
respectively, where $r_f = 85$ Mpc is the limit distance in our sample and $r_0$ is the initial value of the interval 
for analyses to be determined from the Skewness test. 
We expect, for a negligible distance-dependence, that:  ${}^1H \approx {}^2H\approx H_0$. 
In fact, one can verify if $a_1 \ll 1$ is obtained for the ${}^1H$ fit, and if $b_2 \approx {}^2\overline{H}$ is obtained 
for the ${}^2H$ fit, because in these cases the Skewness test confirms a negligible distance-dependence of the data sample, 
and allows to obtain a good measurement of $H_0$. 
Before to perform these calculations, we have to discuss the errors in the distance data from the ALFALFA survey.

\subsection{Distance errors and the Monte Carlo method}\label{error}

One notices that the distance errors presented in the ALFALFA catalogue are not fully realistic \cite{Haynes18}, 
for this we have to generate synthetic errors from a Gaussian distribution that are, on average, 20\% of their catalogued distance values. 
We implement the error propagation on distances when analysing the $V_{i,\text{ref}} / r_i$ distribution, used for the parameter 
analyses in section~\ref{functions}.

The assumption of a 20\% error in the distances is motivated, firstly, by the type of tracer adopted in the analysis. 
As we are dealing with galaxies in the Local Universe, it is known that, on average, the relative error is around 
20\% -- 25\%~\cite{Boruah20}. 
Our second motivation in assuming a large error is due to the lack of knowledge about which distance method was applied for each galaxy in the 
catalogue~(see section \ref{sec2.1} for more details)~\footnote{Note that, in \cite{Jones18}, the authors need distances to obtain 
	the HI mass function, the main goal of the ALFALFA project. 
	However, due to criticisms, the authors performed a series of tests to check whether the distances used are biased or not. 
	In summary, the tests showed that the adoption of different distance calculation methods did not statistically affect the parameters 
	obtained from the HI mass function fitting.}.

\begin{figure}[h]
	\centering
	\includegraphics[scale=0.6]{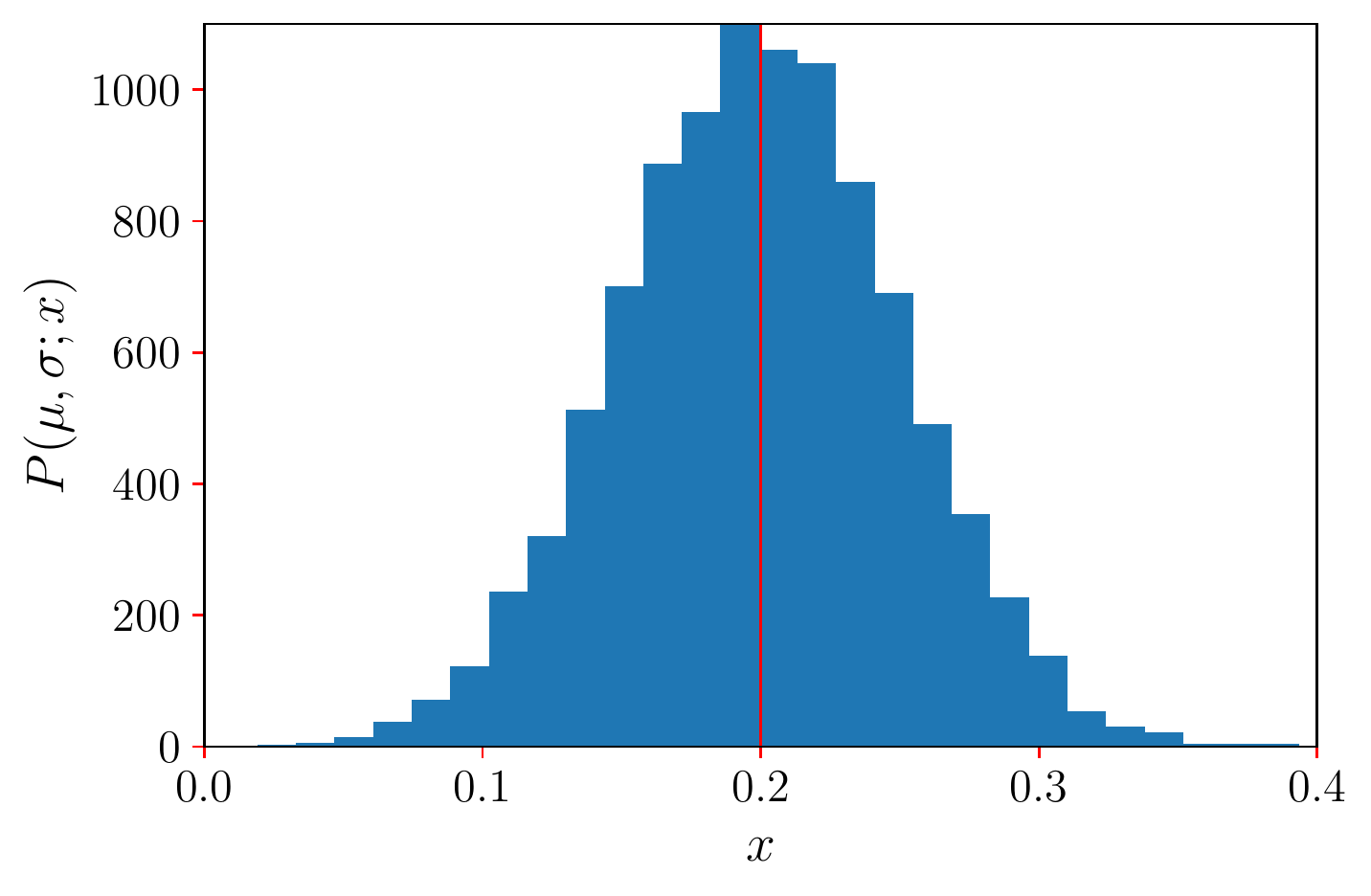}
	\caption{An illustrative example of one of the normal distributions, $P(\mu, \sigma; x)$, generated to obtain the error distribution. 
		For each Monte Carlo realization we generated a new distribution. We fixed $\mu=0.2$ and $\sigma=0.05$.} 
	\label{fig:errordist}
\end{figure}

For a robust analysis, we randomize the error data set using Monte Carlo realizations, that is, 
we performed the fit $N$ times for different error distance distributions. 
For the $j$-th Monte Carlo realization the error on the distance is
\begin{equation}\label{errordistance}
	\sigma_r^j = r \times P(\mu,\sigma; x) \,,
\end{equation}
where
\begin{equation}\label{normaldistribution}
	P(\mu, \sigma; x) = \frac{1}{\sigma\sqrt{2 \pi}}e^{-\frac{(x - \mu)^2}{2 \sigma^2}} \,,
\end{equation}
is a normal distribution with $\sigma=0.05$ and $\mu=0.2$. 
For each Monte Carlo realization we generate a distribution 
$P(\mu,\sigma; x)$~\footnote{\url{https://numpy.org/doc/stable/reference/random/generated/numpy.random.normal.html}}, 
exemplified in Figure~\ref{fig:errordist}.

With the synthetic distance errors, one can propagate them for the distribution $\{ V_{i,\text{ref}} / r_i \}$ 
to adjust the best-fit parameters of the functions ${}^{1}H$ and ${}^{2}H$. 
Thus, we can obtain for all $N$ Monte Carlo realizations that allow to calculate their means 
\begin{equation}\label{MCF1}
	{}^{1}\overline{H}_{\text{MC}} = \frac{1}{N} \sum_j {}^{1}\overline{H}_j + \frac{1}{N}\sum_j \delta {}^{1}\overline{H}_j \,,
\end{equation}
and
\begin{equation}\label{MCF2}
	{}^{2}\overline{H}_{\text{MC}} = \frac{1}{N} \sum_j {}^{2}\overline{H}_j + \frac{1}{N}\sum_j \delta {}^{2}\overline{H}_j \,,
\end{equation}
where
\begin{equation}\label{erroF1}
	\delta {}^{1}\overline{H} = \sqrt{\left(\frac{1}{2}\frac{r_f^2 - r_0^2}{r_f - r_0}\delta a_1\right)^2 + \delta b_1^2} \,,
\end{equation}
and
\begin{equation}\label{erroF2}
	\delta {}^{2}\overline{H} = \sqrt{ \left[\frac{1}{r_f - r_0} \ln \left(\frac{r_f}{r_0} \right)\delta a_2 \right]^2 + \delta b_2^2} \,,
\end{equation}
where $\delta a$ and $\delta b$ are the errors for the adjusted parameters $a_1,a_2$ and $b_1,b_2$.


\section{Results}\label{results}

In this section we present our main results. 
First, we show the results concerning the Skewness test described in section~\ref{skewness}, 
that is, the values of: 
selected intervals, $p$-values, Hubble constants (mean, median, and mode), and the Hubble standard deviation.
Second, we show the results concerning the analyses of distance dependency, section \ref{functions}. 
There, we present the average values of each function, namely, ${}^{1}\overline{H}_{\text{MC}}$ and ${}^{2}\overline{H}_{\text{MC}}$, 
their respective error bars, and the mean values of $a_{\text{MC}}$ and $b_{\text{MC}}$, 
all quantities obtained from Monte Carlo realizations, necessary to implement the error propagation of distances. 
Finally, due to the dipolar behavior observed in the $H_0$ measurements in sections \ref{skewness} and \ref{functions}, we calculate the bulk flow due to peculiar motion in section \ref{bulkflow}.

\subsection{Results of the Skewness test}\label{resultadosskew}

To characterize the statistical properties of the sample in analysis, 
we calculate its mean in three ways: the mean, the median, and the mode, 
and because these values are numerically equal, we are confident that the data 
sample in analysis is (nearly) Gaussian. 
This key result provides us with a measure of data dispersion through the 
standard deviation.
We summarize our main results concerning the Skewness test, and the computation of 
the mean, median, and mode, in the table~\ref{table1}.

As we discussed in the section~\ref{skewness}, for a normal distribution one expects $Z(\sqrt{\beta})\approx 0$. 
We have calculated, for both hemispheres and frames, the value of $Z(\sqrt{\beta})$, using the equation (\ref{zscore}), 
in the interval $20 < r_0 < 70$ Mpc, in steps of 0.5 Mpc. 
For our sample, for both hemispheres, the LG frame of reference presents the best result for $Z$, 
in special the North Hemisphere, with $Z=0.03$. 
However, even with $Z>1$, the $H_0$ values are close to each other, in the CMB frame. 
The worst result of $Z$ presents the lowest $p$-value, around 1\%, which is close to our confidence level.

\begin{figure}
	\centering
	\mbox{    
		\hspace{-0.8cm}
		\includegraphics[scale=0.45]{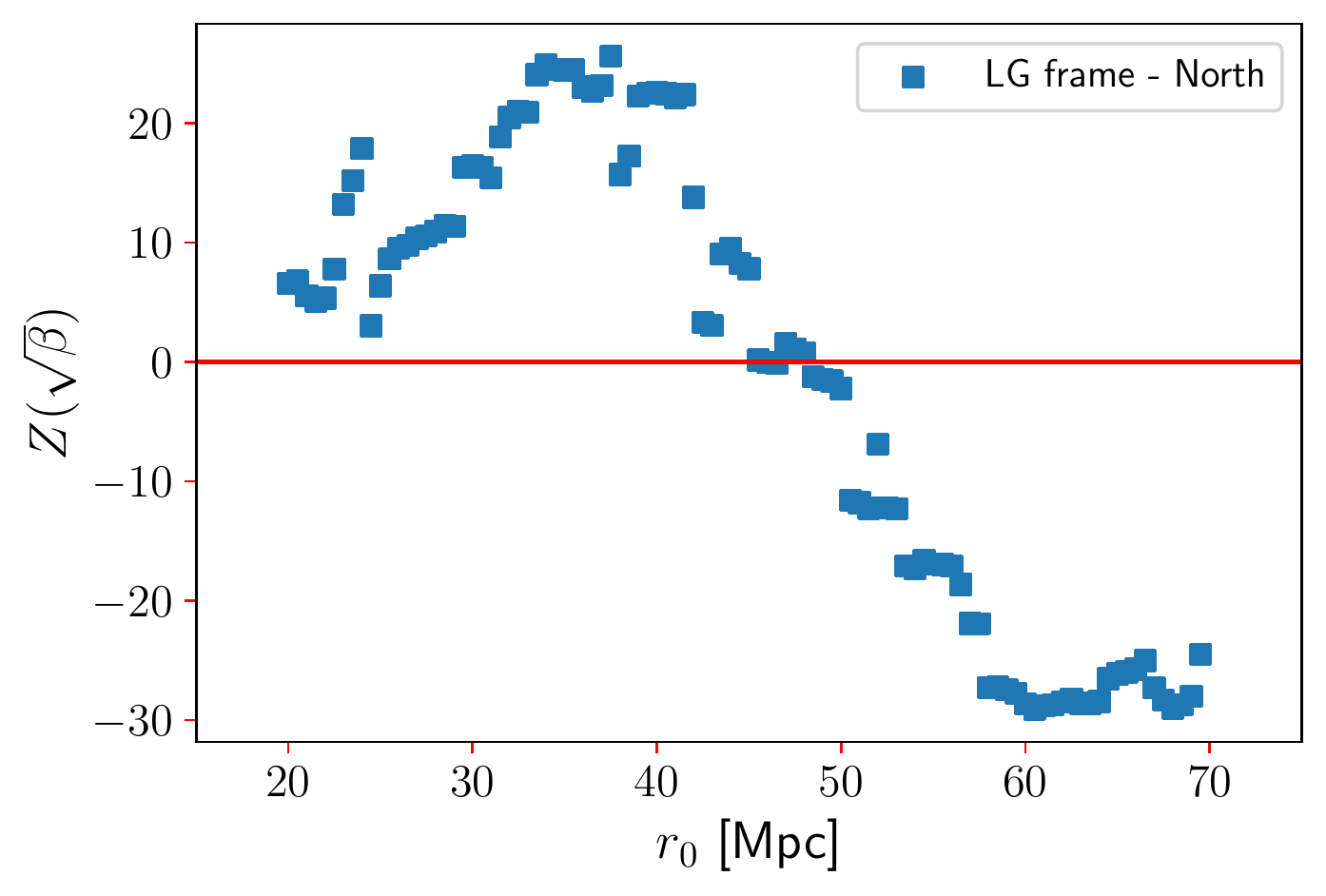}
		\hspace{-0.2cm}    
		\includegraphics[scale=0.45]{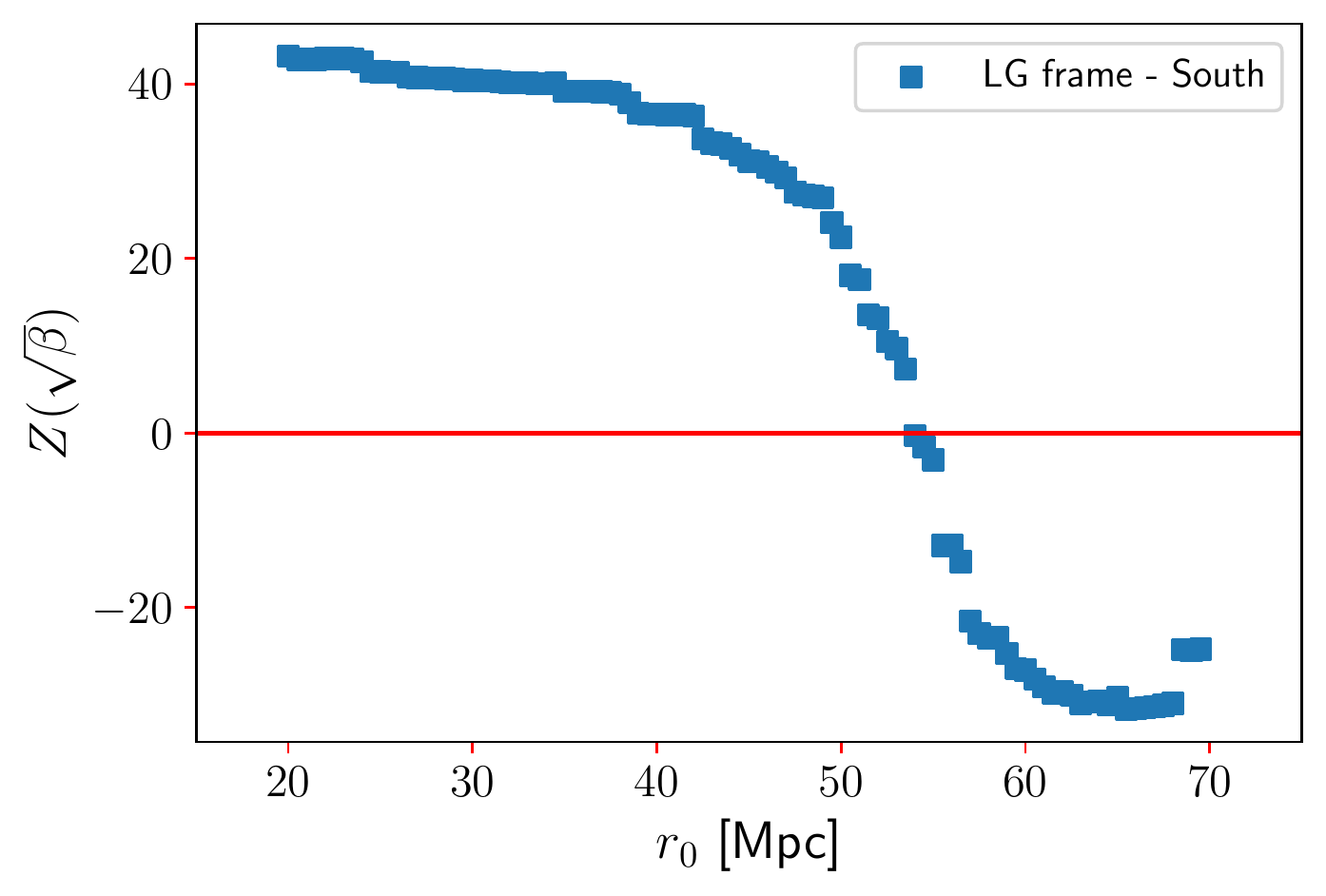}
	}
	\caption{Plots of $Z(\sqrt{\beta})$ versus $r_0$ for the North (left) and South (right) hemispheres, in the LG frame. 
		See the section~\ref{resultadosskew} for details.}
	\label{fig:Z_score_LG}
\end{figure}

\begin{figure}
	\centering
	\mbox{    
		\hspace{-0.8cm}
		\includegraphics[scale=0.45]{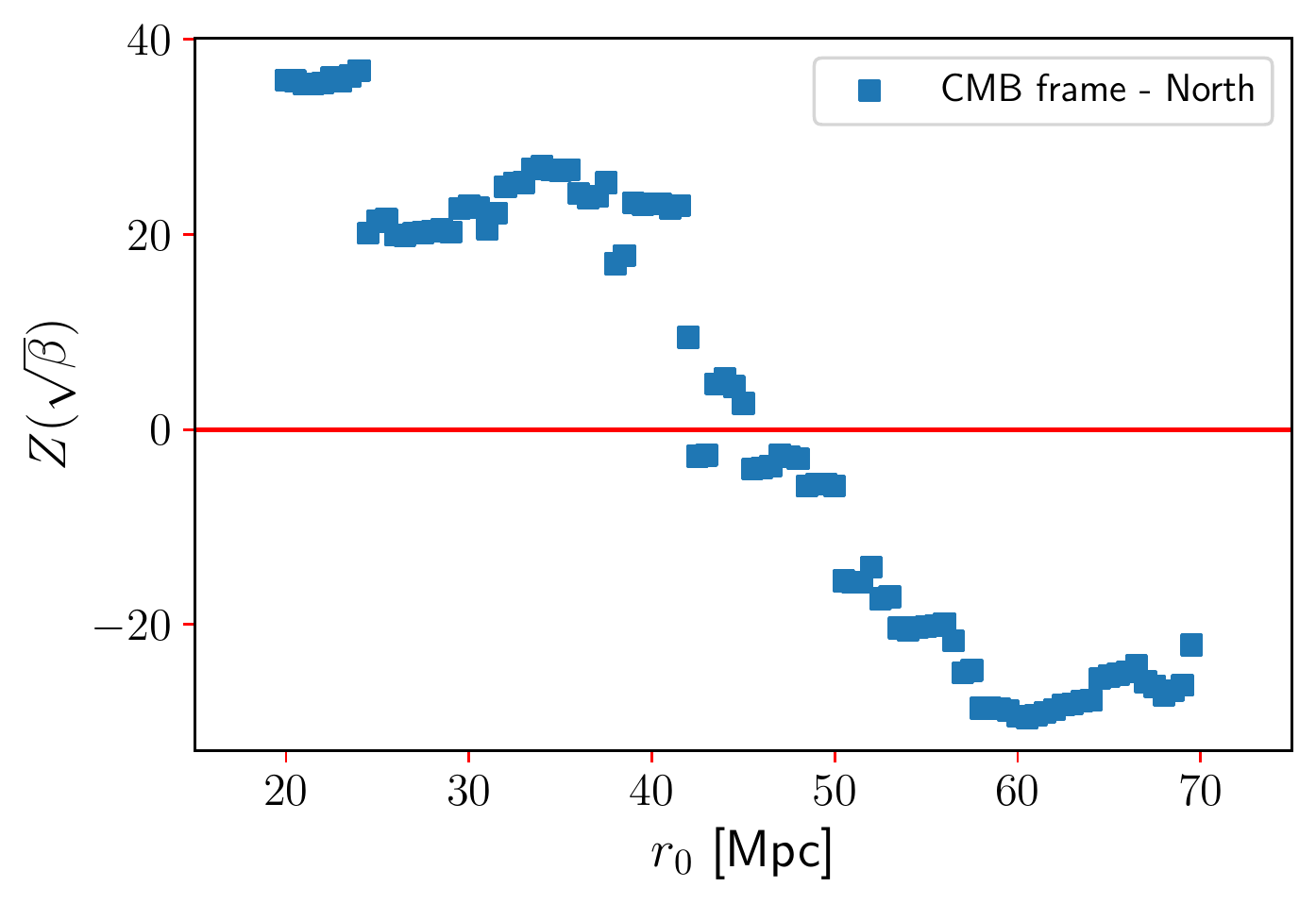}
		\hspace{-0.2cm}
		\includegraphics[scale=0.45]{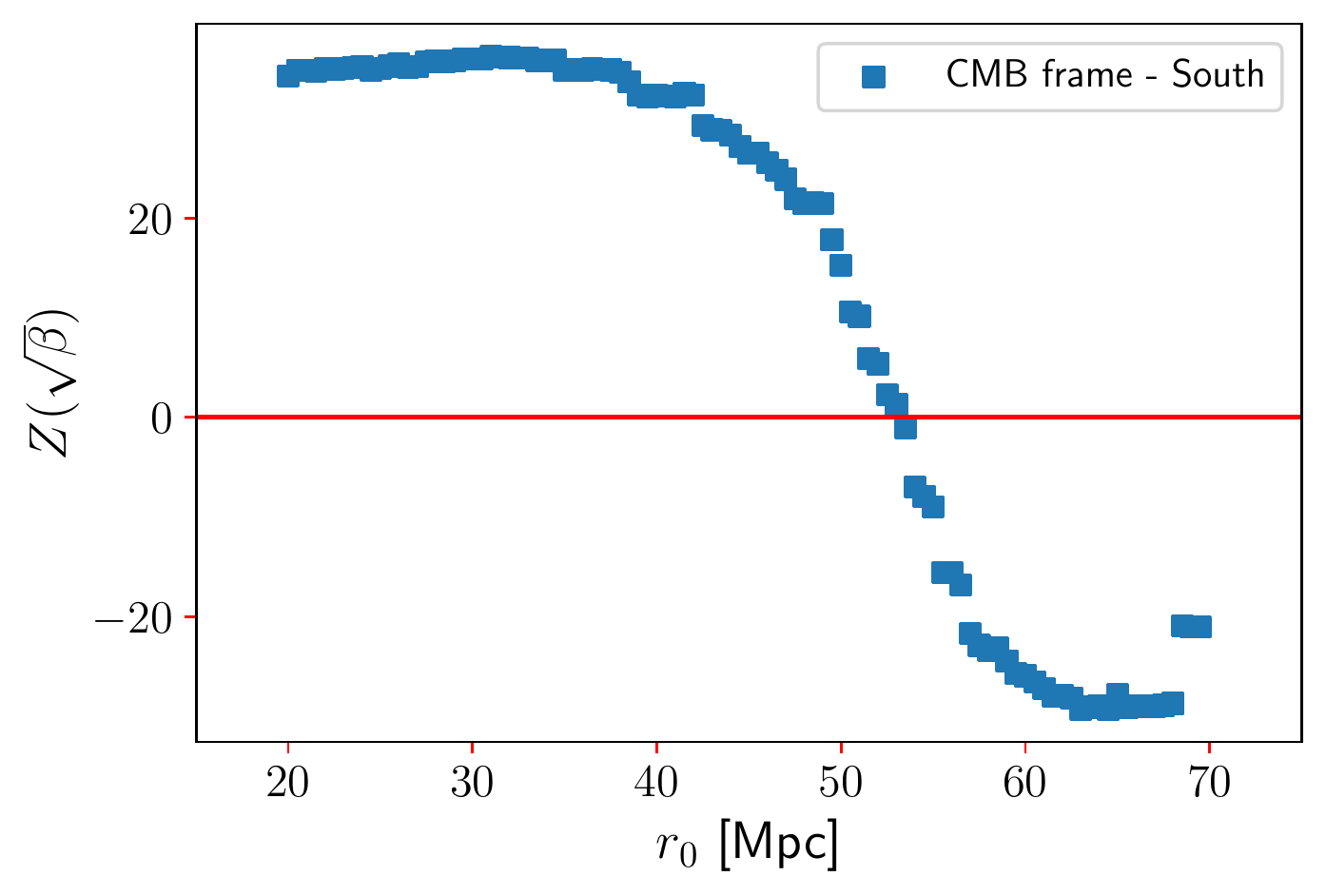}
	}
	\caption{Plots of $Z(\sqrt{\beta})$ versus $r_0$ for the North (left) and South (right) hemispheres, in the CMB frame. 
		See the section~\ref{resultadosskew} for details.}
	\label{fig:Z_score_CMB}
\end{figure}

An interesting feature is noticed when one plots the values of $Z$ as a function of $r_0$, as done in 
Figures~\ref{fig:Z_score_LG} and~\ref{fig:Z_score_CMB} for the LG and CMB frames, respectively. 
In these figures each blue square represents a distribution set that show a normal or Gaussian behavior when 
such square is close to the red line ($Z \approx 0$). 
In both frames these plots exhibit the following feature: 
analyses of the $Z$ function for the South hemisphere show that there are not many intervals, of the type $[r_0, r_f]$, revealing normality;
in contrast in the North hemisphere there are several possible $r_0$ values that define data sets having this property.

For the intervals selected with the Skewness test, we calculate $H_0$ and its respective error, 
using the equations (\ref{mean}) and (\ref{std}), respectively. 
Also, to see if the skewness test selected the best interval, we also calculate the median and mode values;
for the latter, we made the $\{ V_i/r_i \}$ distribution to integer values. 
In the table~\ref{table1}, we present the mode together with the percentage of points with that value. 
For both hemispheres and frames, no discrepancy is observed in the values obtained for the mean, median, and mode. 

Analysing the values of $H_0$, we can notice that, there is a decrease (close to half) in the difference between hemispheres when we go from LG to CMB for the mean result, $\delta H_{\text{LG}} = 8.32\pm5.59$ km/s/Mpc and $\delta H_{\text{CMB}} = 4.35\pm5.78$ km/s/Mpc. Considering the size of the error bars, there is no significant difference between the hemispheres for distances greater than 46 Mpc, thus our sample present an isotropy result for $H_0$. Combining both hemispheres (CMB frame) using a weighted average scheme (see appendix \ref{ApendiceB}, equation (\ref{WAM})), the result for the Hubble-Lema\^itre constant using the Skewness test is $H_0=69.52\pm2.85$ km/s/Mpc, which is in accordance with the Planck results \cite{Planck20} within $1\sigma$ and with the Riess et al. measurement~\cite{Riess21} within $2\sigma$. 

\linespread{1.2}
\begin{table}[h]
	\centering
	\caption{Table with the results of the Skewness, test described in section~\ref{skewness}.}
	\begin{tabular}{|l|ll|ll|}
		\hline
		\multirow{2}{*}{}            & \multicolumn{2}{c|}{North}                               & \multicolumn{2}{c|}{South}                               \\ \cline{2-5} 
		& \multicolumn{1}{l|}{LG}               & CMB              & \multicolumn{1}{l|}{LG}               & CMB              \\ \hline
		$Z(\sqrt{\beta})$            & \multicolumn{1}{l|}{0.03}             & 2.62             & \multicolumn{1}{l|}{0.33}             & 1.08             \\ \hline
		$p$-value                    & \multicolumn{1}{l|}{0.98}             & 0.009            & \multicolumn{1}{l|}{0.74}             & 0.28             \\ \hline
		$r_0$ {[}Mpc{]}              & \multicolumn{1}{l|}{46.0}             & 47.0             & \multicolumn{1}{l|}{54.0}             & 53.5             \\ \hline
		$H_0$ mean {[}km/s/Mpc{]}   & \multicolumn{1}{l|}{66.28 $\pm$ 3.61} & 71.32 $\pm$ 3.72 & \multicolumn{1}{l|}{74.60 $\pm$ 4.27} & 66.97 $\pm$ 4.43 \\ \hline
		$H_0$ median {[}km/s/Mpc{]} & \multicolumn{1}{l|}{66.74}            & 71.89            & \multicolumn{1}{l|}{74.33}            & 66.90            \\ \hline
		$H_0$ mode {[}km/s/Mpc{]}   & \multicolumn{1}{l|}{66 / 26.79\%}     & 72 / 35.72\%     & \multicolumn{1}{l|}{74 / 24.50\%}     & 65 / 13.68\%      \\ \hline
	\end{tabular}
	\label{table1}
\end{table}

\subsection{Results of distance dependence analyses}\label{resultadosfuncao}

In this section, we present the results of the distance dependence analyses (DDA) of the intervals selected with the Skewness test. 
This is necessary for two reasons: Despite the convergence of $\{ V_i/r_i \}$ values observed in the North and South samples, 
we must analyze whether there is a possible dependence with distance in our results, because the Skewness test is not able to detect 
such dependence and we need to implement the distance errors in the analysis to obtain a robust result of $H_0^N$ and $H_0^S$. 

We summarize our results in the table~\ref{table2}, where the analysis is divided by hemispheres and frames of reference. 
The initial values for the interval, $r_0$, were presented in the table~\ref{table1}. 
We used $N = 30,000$ Monte Carlo realizations, which presents a good convergence for the values analysed. 
The values presented in the table \ref{table2} are mean values over the all Monte Carlos realizations. 
For the average values of ${}^{1}\overline{H}$ and ${}^{2}\overline{H}$ and their respective errors we use
equations (\ref{MCF1}), (\ref{MCF2}), (\ref{erroF1}), and (\ref{erroF2}).

Another way to analyze the model fit is to look the values of the adjusted parameters, displayed in table~\ref{table2}. 
For the ${}^{1}\overline{H}$ function was expected $a_{\text{MC}}\ll 1$ and actually we observe $a_{\text{MC}}\ll 1$ 
in both hemispheres and both frames. 
For the ${}^{2}\overline{H}$ function the parameter $b_{\text{MC}}$ was expected to have a value close to $H_0$: 
$b_{\text{MC}} \simeq H_0$, but this is only observed for the North Hemisphere in the CMB frame. 
This result indicates a weak convergence of the data. 
Therefore, we conclude that the ${}^{1}\overline{H}$ function provides the best performance in the calculation of 
the Hubble constant $H_0$ using the DDA methodology. 
Then, combining the $H_0$ values from both hemispheres in the CMB frame, using a weighted average scheme 
(see the equation (\ref{WAM}) in the appendix~\ref{ApendiceB}), we find $H_0 = 69.03 \pm 1.87$ km/s/Mpc, 
a result that is in good agreement with that obtained with the Skewness test (section~\ref{resultadosskew}), 
and moreover also in concordance with the Planck result \cite{Planck20} within $1\sigma$, and with the Riess et al. measurement~\cite{Riess21} within $2\sigma$.

\begin{figure}
	\centering
	\includegraphics[scale=0.4]{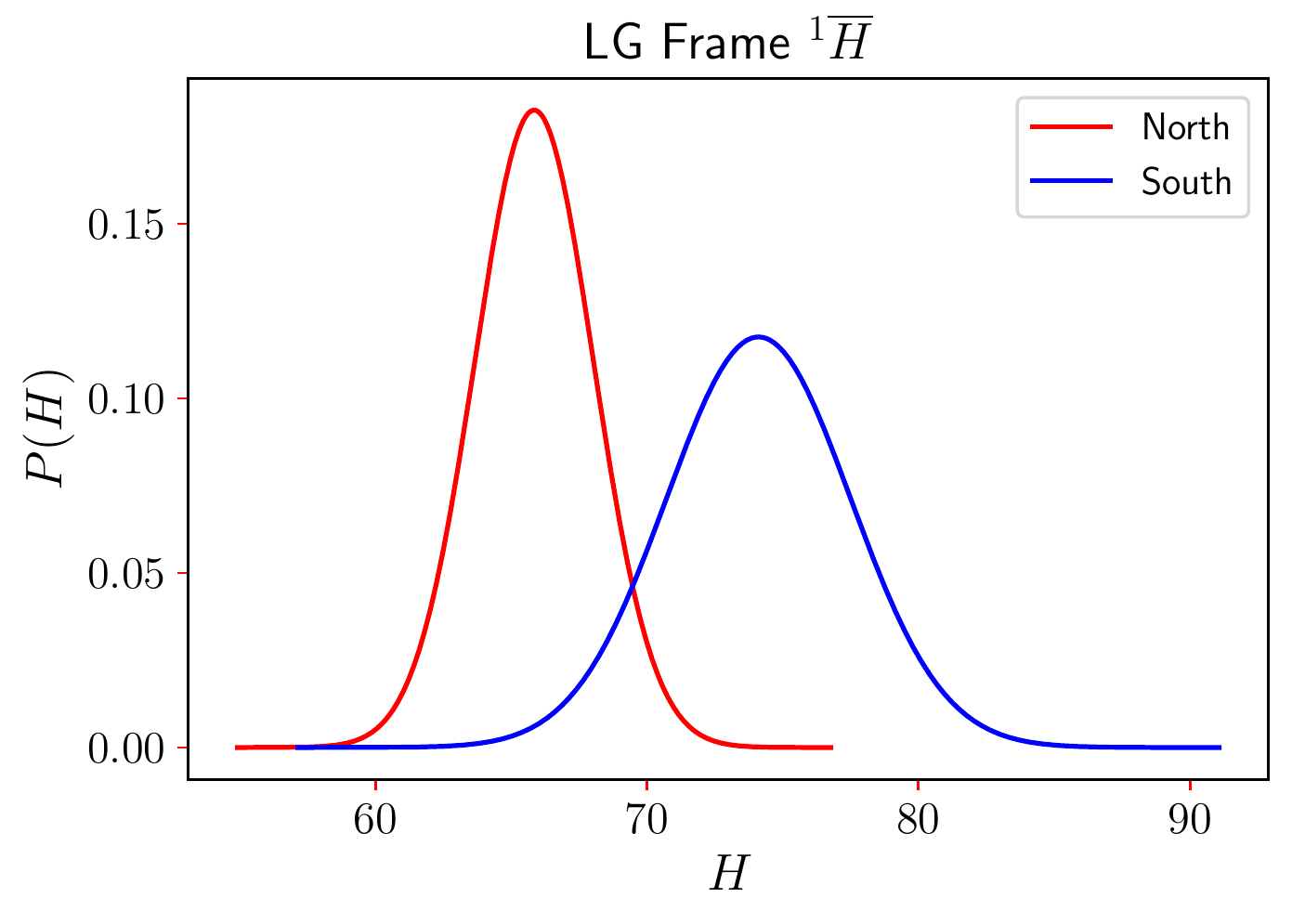}
	\includegraphics[scale=0.4]{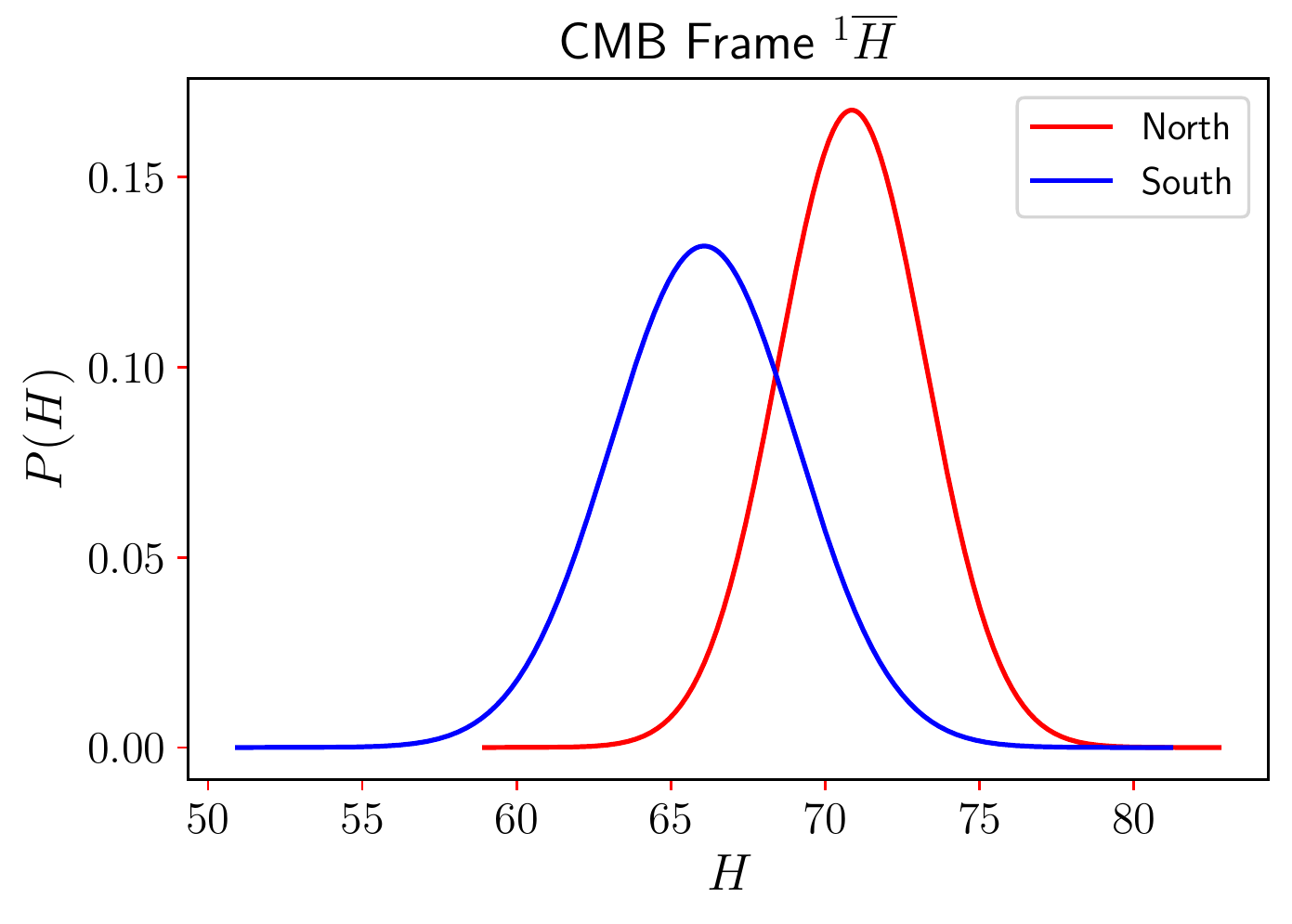}
	\caption{The calculation of the Hubble constant in the LG (left) and CMB (right) frames for the function ${}^{1}\overline{H}$. 
		See the sections~\ref{functions} and~\ref{resultadosfuncao} for details.}
	\label{fig:LG_frame_function}
\end{figure}
\begin{figure}
	\centering
	\includegraphics[scale=0.4]{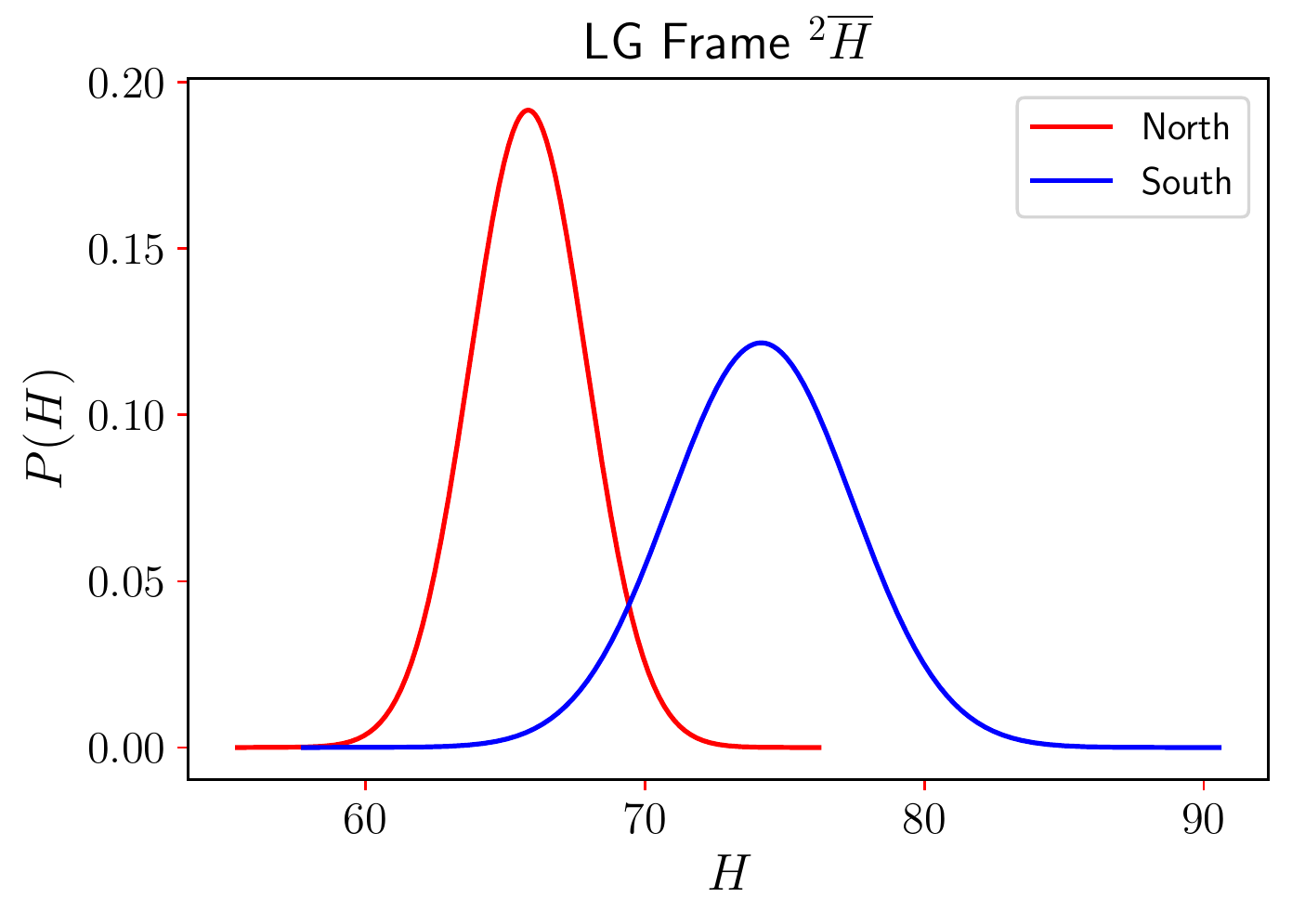}
	\includegraphics[scale=0.4]{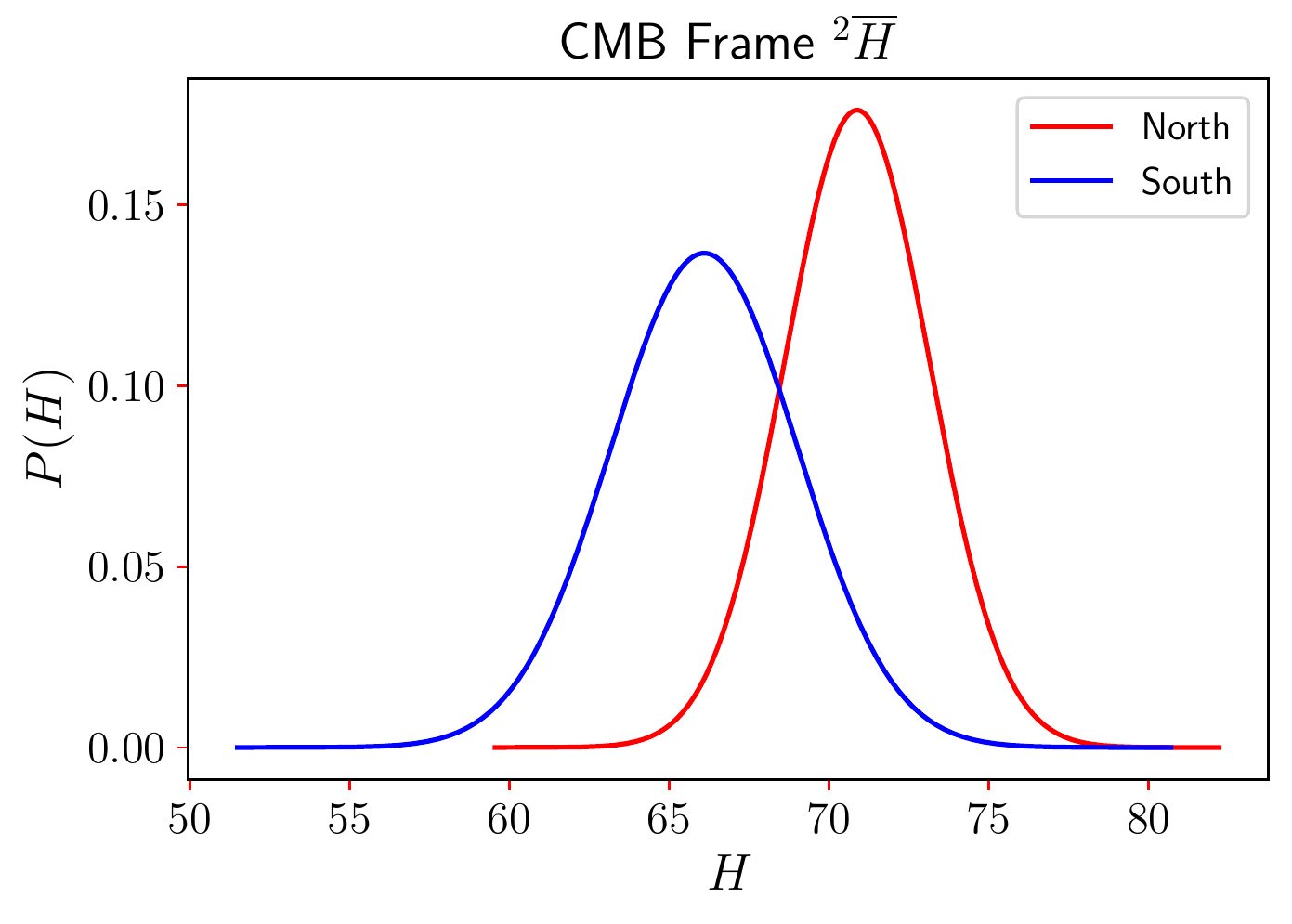}
	\caption{The calculation of the Hubble constant in the LG (left) and CMB (right) frames for the function ${}^{2}\overline{H}$. 
		See the sections~\ref{functions} and~\ref{resultadosfuncao} for details.}
	\label{fig:CMB_frame_function}
\end{figure}

In the Figures~\ref{fig:LG_frame_function}~(${}^{1}\overline{H}$ function) and~\ref{fig:CMB_frame_function}~(${}^{2}\overline{H}$ function) we show the results of $H_0$ for both 
hemispheres and frames of reference. 
We observe a robust result for $H_0$ when we compare both analyses Skewness and DDA, although for the DDA approach 
we have an improvement in the error. 
The difference obtained in antipodal hemispheres remains, basically, the same when compared with the Skewness test: 
$\delta H_{\text{LG}} = 8.26 \pm 4.05$ km/s/Mpc and $\delta H_{\text{CMB}} = 4.80\pm3.85$ km/s/Mpc for ${}^{1}\overline{H}$, 
$\delta H_{\text{LG}} = 8.35\pm3.91$ km/s/Mpc and $\delta H_{\text{CMB}} = 4.77\pm3.69$ km/s/Mpc for ${}^{2}\overline{H}$. 
The fact that we obtain $\delta H_{\text{CMB}} \ne 0$ in both analyses in the CMB frame, for a $1\sigma$ concordance level, is indicative of a residual 
motion not accounted for in the velocity-frame transformations considered, but imprinted in the catalogue in study. 
In fact, we discuss this interesting point in the next section.

\linespread{1.2}
\begin{table}[h]
	\centering
	\caption{Table with the results referring to the distance-dependence analyses, methodology presented in section~\ref{functions}. 
		For the discussion of these results, see section~\ref{resultadosfuncao}.}
	\label{table2}
	\begin{tabular}{|lllll|}
		\hline
		\multicolumn{1}{|l|}{\multirow{2}{*}{}}                  & \multicolumn{2}{c|}{North}                                                    & \multicolumn{2}{c|}{South}                               \\ \cline{2-5} 
		\multicolumn{1}{|l|}{}                                   & \multicolumn{1}{l|}{LG}               & \multicolumn{1}{l|}{CMB}              & \multicolumn{1}{l|}{LG}               & CMB              \\ \hline
		\multicolumn{5}{|c|}{${}^{1}\overline{H}$ function}                                                                                                                                                 \\ \hline
		\multicolumn{1}{|l|}{$H_0$ {[} km/s/Mpc {]}}             & \multicolumn{1}{l|}{65.80 $\pm$ 2.11} & \multicolumn{1}{l|}{70.87 $\pm$ 2.38} & \multicolumn{1}{l|}{74.06 $\pm$ 3.46} & 66.07 $\pm$ 3.02 \\ \hline
		\multicolumn{1}{|l|}{$a_{\text{MC}}$ {[}km/s/Mpc$^2${]}} & \multicolumn{1}{l|}{0.08 $\pm$ 0.02}  & \multicolumn{1}{l|}{$-0.04 \pm$ 0.02} & \multicolumn{1}{l|}{$-0.17 \pm$ 0.03} & $-0.05 \pm$ 0.03 \\ \hline
		\multicolumn{5}{|c|}{${}^{2}\overline{H}$ function}                                                                                                                                                 \\ \hline
		\multicolumn{1}{|l|}{$H_0$ {[}km/s/Mpc{]}}               & \multicolumn{1}{l|}{65.77 $\pm$ 2.01} & \multicolumn{1}{l|}{70.87 $\pm$ 2.26} & \multicolumn{1}{l|}{74.12 $\pm$ 3.35} & 66.10 $\pm$ 2.92 \\ \hline
		\multicolumn{1}{|l|}{$b_{\text{MC}}$ {[}km/s/Mpc{]}}     & \multicolumn{1}{l|}{71.13 $\pm$ 1.42} & \multicolumn{1}{l|}{68.23 $\pm$ 1.60} & \multicolumn{1}{l|}{62.87 $\pm$ 2.36} & 62.58 $\pm$ 2.06 \\ \hline
	\end{tabular}
\end{table}

\newpage
\subsection{Bulk flow velocity with ALFALFA}\label{bulkflow}

%
Our measurements of the Hubble constant in the CMB reference frame in opposite 
hemispheres, $H_0^N$ and $H_0^S$, show a dipolar feature, as observed in 
table~\ref{table2}: $H_0^N = 70.87 \pm 2.38$, $H_0^S = 66.07 \pm 3.02$. 
This difference is attributed to our motion relative to the bulk flow in the 
Local Universe (see, e.g., refs.~\cite{Hong14,Scrimgeour16,Qin19}). 

\begin{figure}[H]
\centering
\includegraphics[scale=0.4]{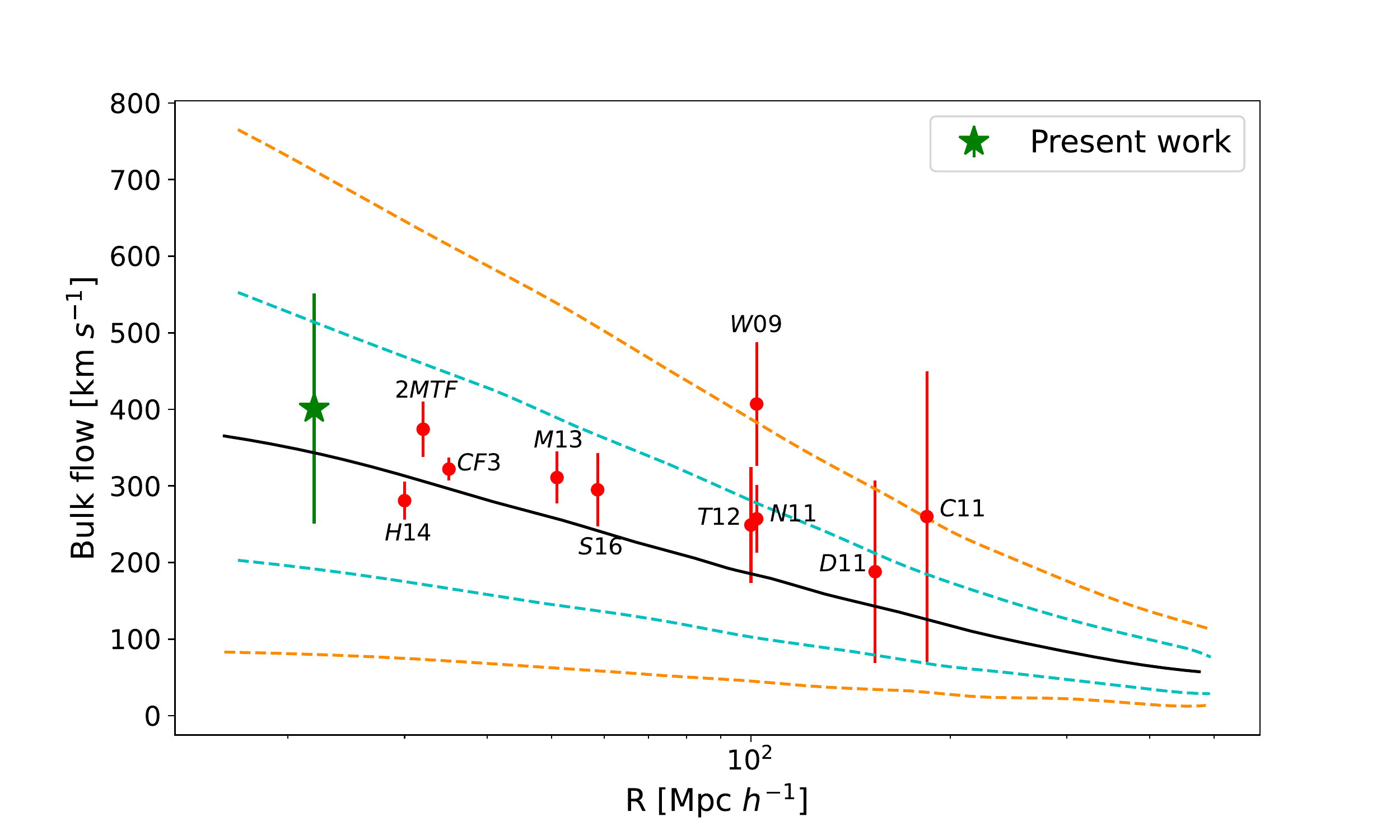}
\caption{Our measurement of the bulk flow velocity (a green star) compared with other values from the literature. The bulk flow velocity expected in the $\Lambda$CDM model is shown as a black line; the light-blue and orange dashed lines indicate $68\%$ and $95\%$ confidence levels, respectively (see, e.g., ref.~\cite{Qin19}).}
\label{fig:bulkflow}
\end{figure}

Considering that $\theta$ is the angle between this dipole, 
$\overrightarrow{\delta H_0} \equiv \overrightarrow{\delta H_0^N} - \overrightarrow{\delta H_0^S}$, 
with the bulk flow direction given in~\cite{Hong14} one finds $\theta \simeq 68^{\circ}$. 
The magnitude of the dipole $\delta H_0$ is proportional to the projection of the bulk flow velocity and inversely proportional 
to the effective distance, $R$, of the sample in analysis (see ref.~\cite{Scrimgeour16}), that is 
\begin{eqnarray}
\delta H_0 = \frac{V_{\mbox{\tiny BF}} \cos \theta}{R} \,,
\end{eqnarray}
where the effective distance $R = 31.3 \pm 6.26$ Mpc is calculated following ref.~\cite{Scrimgeour16} 
according to the features of the ALFALFA survey. 
Then 
\begin{eqnarray}
V_{\mbox{\tiny BF}} &=& \frac{\delta H_0 \, R}{\cos \theta} \,, \\
V_{\mbox{\tiny BF}} &=& \frac{(4.80 \pm 3.85) \, 
(31.3 \pm 6.26)}{0.3746} \,,
\end{eqnarray}
where one finds $V_{\mbox{\tiny BF}} = 401.06 \pm 150.55$ km/s, a result that is in good agreement with 
the literature~\cite{Watkins}, and also with the value expected in the  standard model of cosmology 
(see Figure~\ref{fig:bulkflow} and table~\ref{tableBF}). 
Clearly, one expects that for deeper catalogues the effect of this bulk flow velocity goes to zero, 
and this effect can be seen in Figure~\ref{fig:bulkflow}.

\linespread{1.2}
\begin{table}[h]
\centering
\caption{Table with different bulk flow velocities found in the literature: 
H14~\cite{Hong14}; 2MTF and CF3~\cite{Qin19}; M13~\cite{Ma}; S16~\cite{Scrimgeour16}; T12~\cite{Turnbull}; N11~\cite{Nusser}; W09~\cite{Watkins}; D11~\cite{Dai}, and 
C11~\cite{Colin}. 
Notice that, to put our measurement in the same units as those of the literature, we have used $h=0.7$ for the effective distance.}
\label{tableBF}
\begin{tabular}{|c|c|c|}
\hline
Bulk flow measurements & R [Mpc $h^{-1}]$ & $V_{BF}$ [km $s^{-1}$] \\
\hline 
Present work & 21.91 & 401.06 $\pm$ 150.55\\
\hline
H14 & 30 & 280.8 $\pm$ 25\\
\hline
CF3 & 35 & 322 $\pm$ 15\\
\hline
2MTF & 32 & 374 $\pm$ 36\\
\hline
M13 & 51 & 310.9 $\pm$ 33.9 \\
\hline
S16 & 59 & 295 $\pm$ 48 \\
\hline
T12 & 100 & 249 $\pm$ 76\\
\hline
N11 & 102 & 257 $\pm$ 44\\
\hline
W09 & 102 & 407 $\pm$ 81\\
\hline
D11 & 154 & 188 $\pm$ 119\\
\hline
C11 & 184 & 260 $\pm$ 190 \\
\hline
\end{tabular}
\end{table}

\section{Conclusions and Final remarks}\label{conclusions}

It is important to know the main features of the bulk flow in the Local Universe for a better determination of the relative motions there, an information that would contribute to a precise calculation of the Hubble-Lema\^{\i}tre law at very low redshifts~\cite{Qin19}. 
The current literature reports different analyses to measure the Hubble constant but the common approach is to consider analyses with data $z \gtrsim 0.02$, 
avoiding the effect caused by the peculiar velocities.

In fact, the gravitational field originated by the matter distribution dictates the peculiar motions of the galaxies. 
This is true anywhere in the universe, but it has more severe consequences in the Local Universe where 
this effect competes with the universe expansion, affecting the measurements of the recessional velocities 
expected according to the Hubble-Lema\^{\i}tre law, making 
difficult to obtain a measure of the Hubble constant. 
Our methodology aims to compute $H_0$ in two sky regions --in opposite galactic hemispheres-- 
mapped by the ALFALFA survey and look for dependence with distance, direction, and also test for the effect of reference frame 
changes. 
According to our analyses, the Hubble constant in the CMB frame is $H_0 = 69.03\pm 1.87$ km/s/Mpc, 
a result that is in good agreement with the Planck results~\cite{Planck20} within $1\sigma$ and with the Riess et al. value~\cite{Riess21} within $2\sigma$.

We have selected a sample of HI extra-galactic objects from the catalogue of the ALFALFA survey to perform 
a measurement of the Hubble constant $H_0$ in the Local Universe, i.e., $c z_{\odot} < 6\,000$ km/s. 
Our analyses are performed on two small regions of the celestial sphere located in the Northern and Southern 
galactic hemispheres as shown in Figure~\ref{fig2}. 
The velocity frame transformations done in each region evidence a dipolar feature in the $H_0$ values, a 
phenomenon attributed to our motion relative to the bulk flow in the Local Universe (see, e.g., \cite{Hong14,Scrimgeour16,Qin19}). 
In fact, the computation of $H_0$ in opposite galactic hemispheres, $H_0^N = 70.87 \pm 2.38$ and $H_0^S = 66.07 \pm 3.02$, 
allows us to measure the bulk flow velocity $V_{\mbox{\tiny BF}} = 401.06 \pm 150.55$ km/s at the effective distance 
$31.3 \pm 6.26$ Mpc, a novel result found analysing the ALFALFA data (our analyses and results are shown in section~\ref{bulkflow}). 
We then confirm the influence of the {\em bulk flow} on the structures of the Local Universe which manifests through a dipolar 
behavior of the Hubble constant in opposite hemispheres.


\bmhead{Acknowledgments}

FA, JO, MLSD, and AB thank CAPES and CNPq for the grants under which this work was carried out.

\section*{Declarations}
\textbf{Competing Interest} The authors declare that they have no competing interest.
\\[10pt]
This version of the article has been accepted for publication after peer review  but is not the Version of Record and does not reflect post-acceptance improvements, or any corrections. The Version of Record is available online at: \url{https://doi.org/10.1007/s13538-023-01259-z}. Use of this Accepted Version is subject to the publisher’s Accepted Manuscript terms of use \url{https://www.springernature.com/gp/open-research/policies/accepted-manuscript-terms}.

\begin{appendices}

\section{Null test for Skewness test}\label{ApendiceA}

In this work we propose to use the skewness test~\cite{Agostino90} to select the best interval for our analyses. 
For this, we use the public code SciPy~\footnote{\url{https://docs.scipy.org/doc/scipy/index.html}} which has the 
Skewness test  library~\footnote{\url{https://docs.scipy.org/doc/scipy/reference/generated/scipy.stats.skewtest.html}}.

Firstly, we develop a simple null test to see if the code correctly returns the values of Z and $p$-values. 
We create a normal distribution with $N=10,000$ objects, and $(\mu, \sigma) = (0, 1)$. 
Notice that one does not obtain $Z \simeq 0$ for a single run, as a matter of fact, one needs to run the numerical 
code ${\cal N}$ times to appreciate the results.

\begin{figure}[h!]
	\centering
	\includegraphics[scale=0.4]{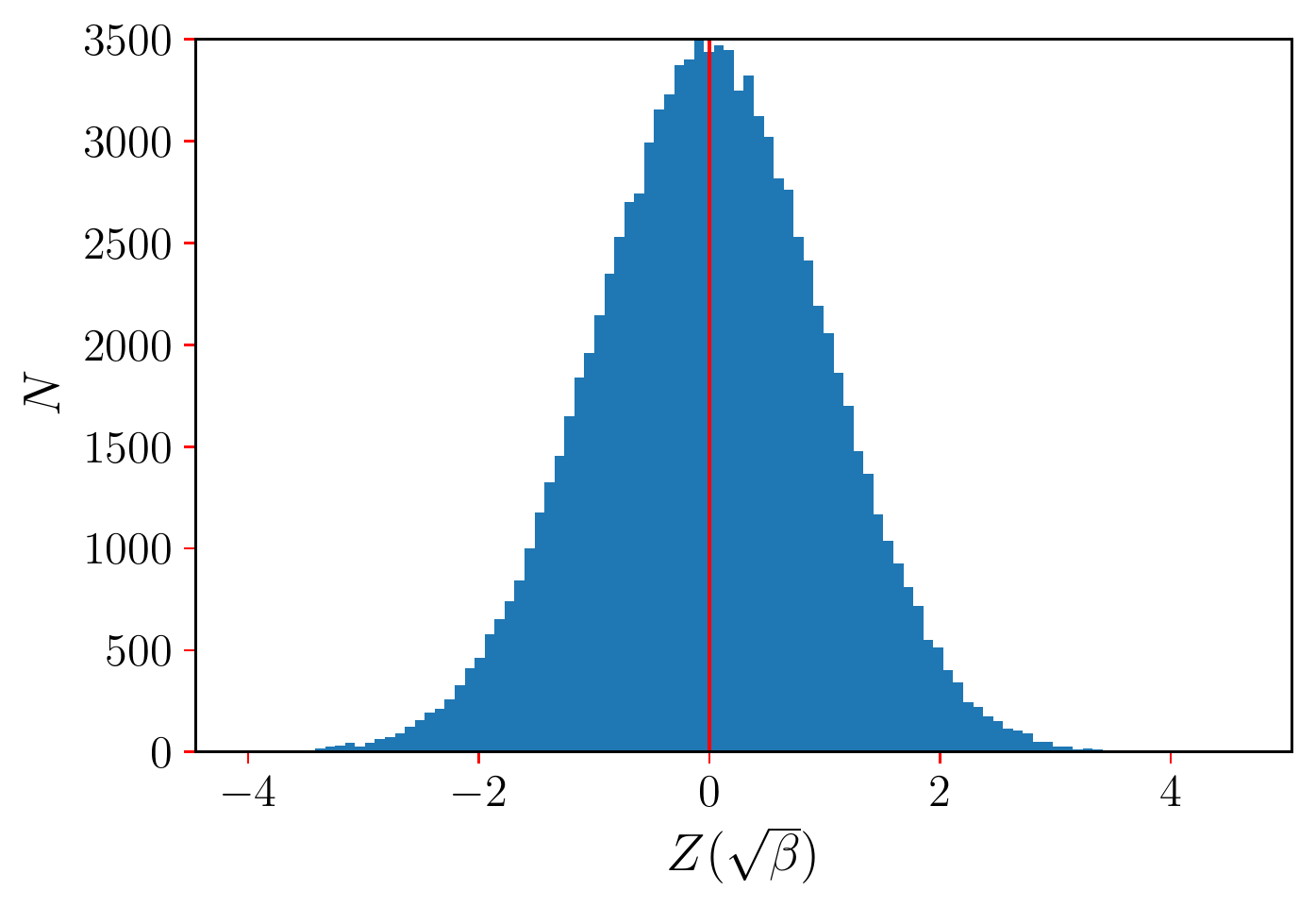}
	\includegraphics[scale=0.4]{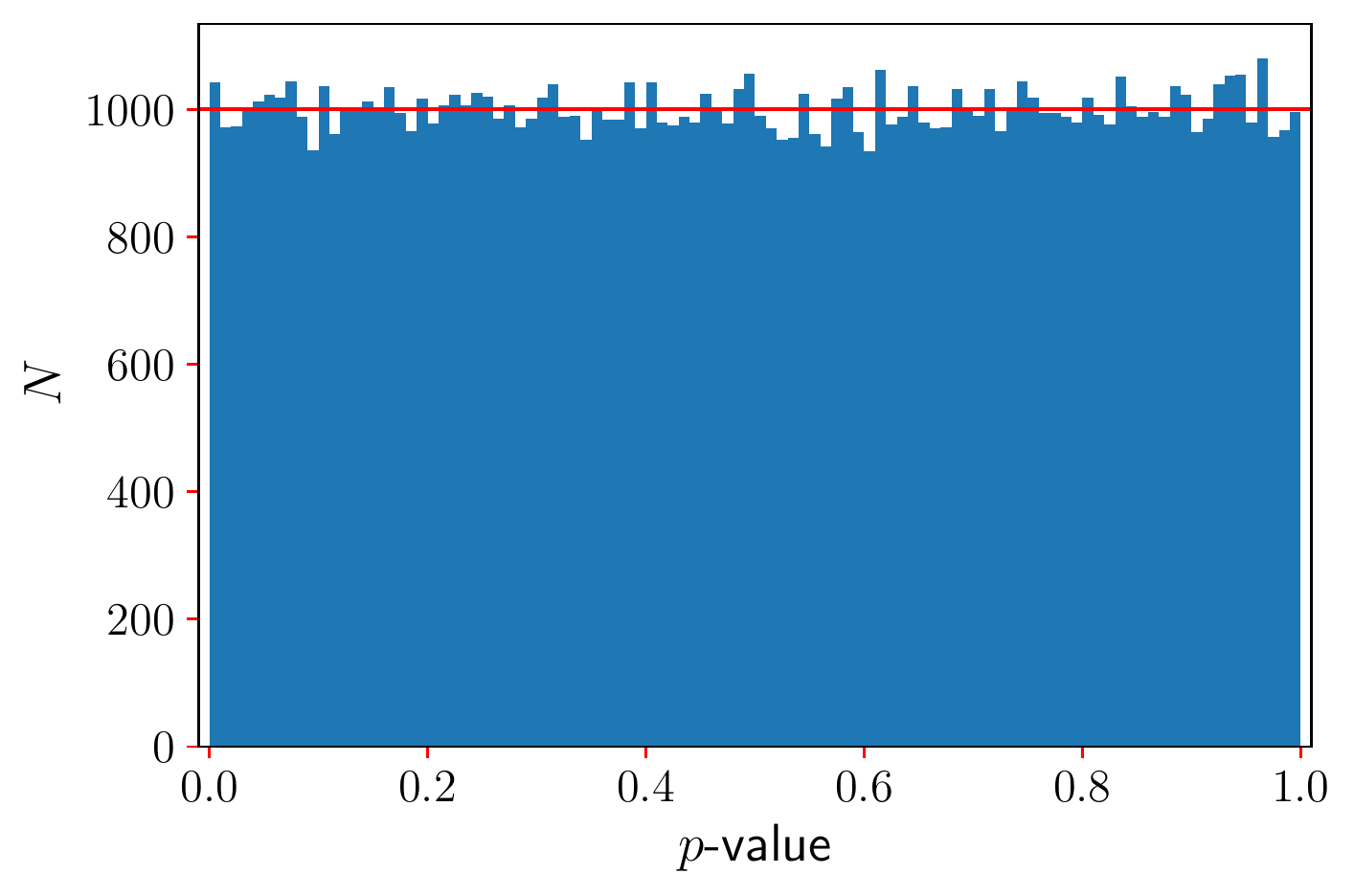}
	\caption{Results of the null test for the code used in this work to perform the Skewness test. 
		Left: Histogram of the $Z$ values calculated for the 100,000 normal distributions. 
		Right: Uniform distribution for the $p$-values obtained from these 100,000 normal distributions.
}
	\label{fig:null_test}
\end{figure}

We run our code ${\cal N} = 100,000$ times and present our results of this null test in the Figure~\ref{fig:null_test}. 
In the left panel we observe the histogram of the skewness values $Z$, for the ${\cal N}$ runs, 
where the mean value is 0, and the variance is 1, as expected. 
In the right panel, we have an uniform distribution for the $p$-values, which also is expected, as for each value of $Z$, 
where the $p$-value varies uniformly between 0 and 1, i.e., without preference for any value.

%

\section{Robustness analyses with several mean measurements}\label{ApendiceB}

In addition to the Skewness test that determines the best interval for our analyses, we would like to examine 
if other intervals produce similar results for $H_0$. 
One way to do this is performing the average procedures with diverse statistical tools. 
Here we shall analyze our data samples with the weighted {\em arithmetic}, {\em geometric}, and {\em harmonic} 
means to study the behavior of the observable $H_0$ in the data sample after performing the LG and CMB transformations.

\subsection{Arithmetic, Geometric and Harmonic means}

We define the \textit{Weighted Arithmetic Mean} (WAM) as
\begin{equation}\label{WAM}
\overline{H}_0^{\text{\tiny \,WAM}} \equiv \frac{ \sum_i H_{i} w_i}{\sum_i w_i} \,,
\end{equation}
where $H_{i} = V_{i, ref.}/r$ which depends on the frame of reference to which the velocities 
have been transformed; $w_i$ are the weights associated with the $n$ measurements and are defined by 
\begin{equation}\label{Weigh weighted average}
w_i = \frac{1}{\sigma_i^2} \,,
\end{equation}
$\sigma_i$ are simulated errors defined in equation (\ref{errordistance}) for $H_i$. 
The error associated with the WAM is
\begin{equation}
\sigma^{\text{\tiny \,WAM}} = \frac{1}{\sqrt{\sum_i w_i}} \,.
\end{equation}

Next, we define the \textit{Weighted Geometric Mean} (WGM) as
\begin{equation}\label{WGM}
\overline{H}_0^{\text{\tiny \,WGM}} \equiv \exp\left(\dfrac{\sum_i w_i\ln(H_i)}{\sum_i w_i}\right) \,, 
\end{equation}
with standard deviation
\begin{equation}
\sigma^{\text{\tiny \,WGM}} = \sqrt{\frac{(\overline{H}_0^{\text{WGM}})^2}{(\sum_i w_i)^2}\sum_i\frac{w_i}{H_i^2}} \,.
\end{equation}

Lastly, we define the \textit{Weighted Harmonic Mean} (WHM) as
\begin{equation}\label{WHM}
\overline{H}_{0}^{\text{\tiny \,WHM}} \equiv \dfrac{\sum_i w_i}{\sum_i w_i/H_i} \,,    
\end{equation}
with standard deviation
\begin{equation}
\sigma^{\text{\tiny \,WHM}} = \sqrt{\dfrac{(\overline{H}_{0}^{\text{\tiny \,WHM}})^2}{(\sum_i w_i / H_i)^2}\sum_i\frac{w_i}{H_i^4}} \,.
\end{equation}

\subsection{Results}

Our analysis was realized in a sample with $N = 10,000$ for each mean measurement, according to the details described in section~\ref{error}, new analyses that need to be related to the unrealistic errors of the ALFALFA catalog as shown in the equation(~\ref{errordistance}). 
With the appropriate frame transformations, for LG and CMB, this allows us to find $\{ H_i \}$, which are estimates without considering the errors in the  simulated distances and essential to find the mean values. 
Through these simulations, the mean for each hemisphere was extracted for each distance $r_0$ from $40$ Mpc to $85$ Mpc. 
Thus, for each interval a sample of $N$ mean values was generated and to select the value of $H_0$ of this sample, a simple mean was used. 
The error associated with these averages allowed us to calculate the error for the $H_0$ measurements using the same methodology mentioned previously. 

We summarize our results for the WAM, WGM, and WHM in tables~\ref{table3}, \ref{table4}, and \ref{table5}, respectively. 
Note that, for the three means, the results are consistent between each other. 
Moreover, there is a good agreement between this analysis and the Skewness test results presented in table~\ref{table1}.

In these analyses we have a large reduction in the error bar due to the nature of the uncertainty in each mean, since $\sigma \propto 1/\sqrt{n}$, where $n$ is the number of data used to perform the mean. 
However, we can only consider these results as a test because it does not contain the natural dispersion of the data set, as it does in the Skewness test. 
Also, we can not combine in quadrature the dispersion and the mean error because these intervals are not Gaussian or normal distributed, 
as one can verify in the Figures~\ref{fig:Z_score_LG} and~\ref{fig:Z_score_CMB}. 

\linespread{1.2}
\begin{table}[h]
	\centering
	\caption{Table with the results for $H_0$ and $\sigma$ in {[}km/s/Mpc{]} for the WAM.}
	\label{table3}
	\begin{tabular}{|c|ll|ll|}
		\hline
		\multicolumn{1}{|l|}{}                  & \multicolumn{2}{c|}{North}                                    & \multicolumn{2}{c|}{South}                                         \\ \hline
		\multicolumn{1}{|l|}{$r_0$ {[} Mpc {]}} & \multicolumn{1}{c|}{LG}                 & \multicolumn{1}{c|}{CMB} & \multicolumn{1}{c|}{LG}    & \multicolumn{1}{c|}{CMB} \\ \hline
		$40 - 85 $                              & \multicolumn{1}{l|}{65.41 $\pm$ 0.21} & 70.81 $\pm$ 0.23       & \multicolumn{1}{l|}{74.51 $\pm$ 0.23} & 66.10 $\pm$ 0.21       \\ \hline
		$50 - 85$                               & \multicolumn{1}{l|}{65.96 $\pm$ 0.23} & 70.79 $\pm$ 0.25       & \multicolumn{1}{l|}{74.14 $\pm$ 0.24} & 65.10 $\pm$ 0.22       \\ \hline
		$60 - 85$                               & \multicolumn{1}{l|}{66.32 $\pm$ 0.27} & 70.55 $\pm$ 0.29       & \multicolumn{1}{l|}{73.63 $\pm$ 0.27} & 65.91 $\pm$ 0.25       \\ \hline
		$70 - 85$                               & \multicolumn{1}{l|}{66.55 $\pm$ 0.37} & 70.48 $\pm$ 0.40       & \multicolumn{1}{l|}{73.22 $\pm$ 0.35} & 66.11 $\pm$ 0.32       \\ \hline
	\end{tabular}
\end{table}




\linespread{1.2}
\begin{table}[h]
	\centering
	\caption{Table with the results for $H_0$ and $\sigma$ in {[}km/s/Mpc{]} for the WGM.}
	\label{table4}
	\begin{tabular}{|c|ll|ll|}
		\hline
		\multicolumn{1}{|l|}{}                  & \multicolumn{2}{c|}{North}                                       & \multicolumn{2}{c|}{South}                                       \\ \hline
		\multicolumn{1}{|l|}{$r_0$ {[} Mpc {]}} & \multicolumn{1}{c|}{LG}               & \multicolumn{1}{c|}{CMB} & \multicolumn{1}{c|}{LG}               & \multicolumn{1}{c|}{CMB} \\ \hline
		$40 - 85 $                              & \multicolumn{1}{l|}{65.14$\pm$ 0.57} & 70.57 $\pm$ 0.62       & \multicolumn{1}{l|}{74.18 $\pm$ 0.39 } & 65.52 $\pm$  0.31         \\ \hline
		$50 - 85$                               & \multicolumn{1}{l|}{65.80 $\pm$ 0.63} & 70.61 $\pm$ 0.67        & \multicolumn{1}{l|}{73.85 $\pm$ 0.44} & 65.55 $\pm$ 0.35      \\ \hline
		$60 - 85$                               & \multicolumn{1}{l|}{66.17 $\pm$ 0.73} & 70.36 $\pm$ 0.77         & \multicolumn{1}{l|}{73.34 $\pm$ 0.53} & 65.32 $\pm$ 0.42       \\ \hline
		$70 - 85$                               & \multicolumn{1}{l|}{66.47 $\pm$ 0.99} & 70.39 $\pm$ 1.04         & \multicolumn{1}{l|}{73.11 $\pm$ 0.87} & 65.97 $\pm$ 0.71        \\ \hline
	\end{tabular}
\end{table}




\linespread{1.2}
\begin{table}[h]
	\centering
	\caption{Table with the results for $H_0$ and $\sigma$ in {[}km/s/Mpc{]} for the WHM.}
	\label{table5}
	\begin{tabular}{|c|ll|ll|}
		\hline
		\multicolumn{1}{|l|}{}                  & \multicolumn{2}{c|}{North}                                         & \multicolumn{2}{c|}{South}                                         \\ \hline
		\multicolumn{1}{|l|}{$r_0$ {[} Mpc {]}} & \multicolumn{1}{c|}{LG}                 & \multicolumn{1}{c|}{CMB} & \multicolumn{1}{c|}{LG}                 & \multicolumn{1}{c|}{CMB} \\ \hline
		$40 - 85 $                              & \multicolumn{1}{l|}{64.87 $\pm$ 0.23} & 70.36 $\pm$ 0.24      & \multicolumn{1}{l|}{73.84 $\pm$ 0.26} & 64.75 $\pm$ 0.32      \\ \hline
		$50 - 85$                               & \multicolumn{1}{l|}{65.63 $\pm$ 0.24} & 70.44 $\pm$ 0.26      & \multicolumn{1}{l|}{73.51 $\pm$ 0.28} & 64.74 $\pm$ 0.36       \\ \hline
		$60 - 85$                               & \multicolumn{1}{l|}{66.01 $\pm$ 0.28} & 70.17 $\pm$ 0.30     & \multicolumn{1}{l|}{72.98 $\pm$ 0.31} & 64.35 $\pm$ 0.42       \\ \hline
		$70 - 85$                               & \multicolumn{1}{l|}{66.40 $\pm$ 0.38} & 70.31 $\pm$ 0.40       & \multicolumn{1}{l|}{73.00 $\pm$ 0.36} & 65.84 $\pm$ 0.32       \\ \hline
	\end{tabular}
\end{table}



\end{appendices}

\newpage
\bibliographystyle{plain}

\bibliography{sn-article}



\end{document}